%% file: main.tex
\documentclass[11pt,a4paper]{article}
\pdfoutput = 1

\usepackage{jcappub}
\usepackage[utf8]{inputenc}
\usepackage{amsmath,amssymb}
\usepackage{graphicx}
\usepackage{hyperref}
\usepackage{natbib}
\usepackage{xcolor}
\usepackage{physics}
\usepackage{derivative}
\usepackage{subcaption}

\newcommand{\edit}[1]{{#1}}

\newcommand{\bq}{\textbf{q}}
\newcommand{\bp}{\textbf{p}}
\newcommand{\bx}{\textbf{x}}
\newcommand{\bk}{\textbf{k}}

\newcommand{\br}{\textbf{r}}
\newcommand{\bPsi}{\boldsymbol{\Psi}}
\newcommand{\bs}{\textbf{s}}

\newcommand{\half}{\frac{1}{2}}

\def \dtq{\int d^3 \bq \ }

\newcommand{\avg}[1]{\ensuremath{\left\langle #1 \right\rangle}}

\def\Mpc{\, h^{-1} \, {\rm Mpc}}
\def\Mpccube{\, h^{-3} \, {\rm Mpc}^3}

\def\kMpc{\, h \, {\rm Mpc}^{-1}}

\def\PL{P_{\rm lin}}

\title{Cosmological Analysis of Three-Dimensional BOSS Galaxy Clustering and Planck CMB Lensing Cross Correlations via Lagrangian Perturbation Theory }

\author[a]{Shi-Fan Chen}
\author[a,b]{Martin White}
\author[b]{Joseph DeRose}
\author[c,d]{Nickolas Kokron}

\affiliation[a]{Department of Physics, University of California, Berkeley, CA, USA}
\affiliation[b]{Physics Division, Lawrence Berkeley National Laboratory, Berkeley, CA, USA}
\affiliation[c]{Kavli Institute for Particle Astrophysics and Cosmology and Department of Physics, Stanford University, Stanford, CA, USA}
\affiliation[d]{Kavli Institute for Particle Astrophysics and Cosmology, SLAC National Accelerator Laboratory, Menlo Park, CA, USA}

\abstract{We present a formalism for jointly fitting pre- and post-reconstruction redshift-space clustering (RSD) and baryon acoustic oscillations (BAO) plus gravitational lensing (of the CMB) that works directly with the observed 2-point statistics.  The formalism is based upon (effective) Lagrangian perturbation theory and a Lagrangian bias expansion, which models RSD, BAO and galaxy-lensing cross correlations within a consistent dynamical framework. As an example we present an analysis of clustering measured by the Baryon Oscillation Spectroscopic Survey in combination with CMB lensing measured by Planck.  The post-reconstruction BAO strongly constrains the distance-redshift relation, the full-shape redshift-space clustering constrains the matter density and growth rate, and CMB lensing constrains the clustering amplitude.  Using only the redshift space data we obtain $\Omega_\mathrm{m} = 0.303\pm 0.008$, $H_0 = 69.21\pm 0.78$ and $\sigma_8 = 0.743\pm 0.043$.  The addition of lensing information, even when restricted to the Northern Galactic Cap, improves constraints to $\Omega_m = 0.300 \pm 0.008$, $H_0 = 69.21 \pm 0.77$ and $\sigma_8 = 0.707 \pm 0.035$, in tension with CMB and cosmic shear constraints. The combination of $\Omega_m$ and $H_0$ are consistent with Planck, though their constraints derive mostly from redshift-space clustering. The low $\sigma_8$ value are driven by cross correlations with CMB lensing in the low redshift bin ($z\simeq 0.38$) and at large angular scales, which show a $20\%$ deficit compared to expectations from galaxy clustering alone. We conduct several systematics tests on the data and find none that could fully explain these tensions. }

\begin{document}

\maketitle

\section{Introduction}
\label{sec:intro}

The large-scale structure of the Universe provides information on galaxy formation, cosmology and fundamental physics \cite{Pea99,Dod03}.  Perhaps the most powerful measure to date has been the redshift-space two-point function measured by galaxy redshift surveys, which measures both the shape of the primordial power spectrum including distance information in baryon acoustic oscillations (BAO; \cite{Wei13,EH1998,Meiksin99}) and cosmological velocities in the form of redshift-space distortions (RSD; \cite{Kaiser87,Hamilton92}).  A complementary view of large-scale structure comes from gravitational lensing, which probes the projected (Weyl) potential sourced by fluctuations in the matter density.  Of particular interest to us here is the lensing of the cosmic microwave background (CMB) anisotropies, which provide a well-characterized source screen at a well-known redshift \cite{PlanckLegacy18} far behind the lensing potentials.  Either of these probes, or their combination, can be used to measure the amplitude and growth rate of large-scale structure over cosmic time with high precision, providing valuable constraints on our cosmological model and its constituents.

The theoretical study of large scale structure is by now quite mature thanks to continued developments in perturbation theory (PT). Within PT the growth of structure is treated systematically,  order-by-order in the initial conditions with nonlinearities at small scales marginalized away using effective-theory techniques \cite{BNSZ12,CHS12,VlaWhiAvi15}. Biased tracers of large-scale structure like galaxies can similarly be treated by identifying contributions to their clustering at each order allowed by fundamental symmetries \cite{McDRoy09,Sen14,Des16,VlaCasWhi16,Chen20,Fujita+:2020}. Much of this modeling effort has focused on the clustering of galaxies in redshift-space, as measured in spectroscopic surveys, leading to models with accuracy well beyond the expected statistical uncertainty in any realistic surveys \cite{Nishimichi2020,Chen21,Ivanov22}, and which have been tested extensively on existing surveys like BOSS and eBOSS \cite{Ivanov20BOSS,DAmico20,Ivanov:2021zmi,Chen22,Zhang21b,Philcox22}. The same models can also be used to predict weak-lensing measurements, particularly their cross-correlations with galaxy surveys (which allow for cleaner separation of scales by virture of being more localized in redshift). These measurements probe matter clustering and its cross-correlation with galaxy densities, both without redshift-space distortions, and are in fact easier to model within PT since they do not involve large contributions from small-scale velocities. Such predictions have received less attention to date, though there is a long history of PT-inspired models of galaxy lensing cross correlations applied to both simulations and data (e.g.\ refs.~\cite{Baldauf10,Krolewski21,Pandey21}), and full PT models have recently been successfully applied to cross-correlations between Planck CMB lensing and galaxies from the DESI Legacy Imaging Survey \cite{Kitanidis21,White22}.

\begin{figure}
    \centering
    \resizebox{\columnwidth}{!}{\includegraphics{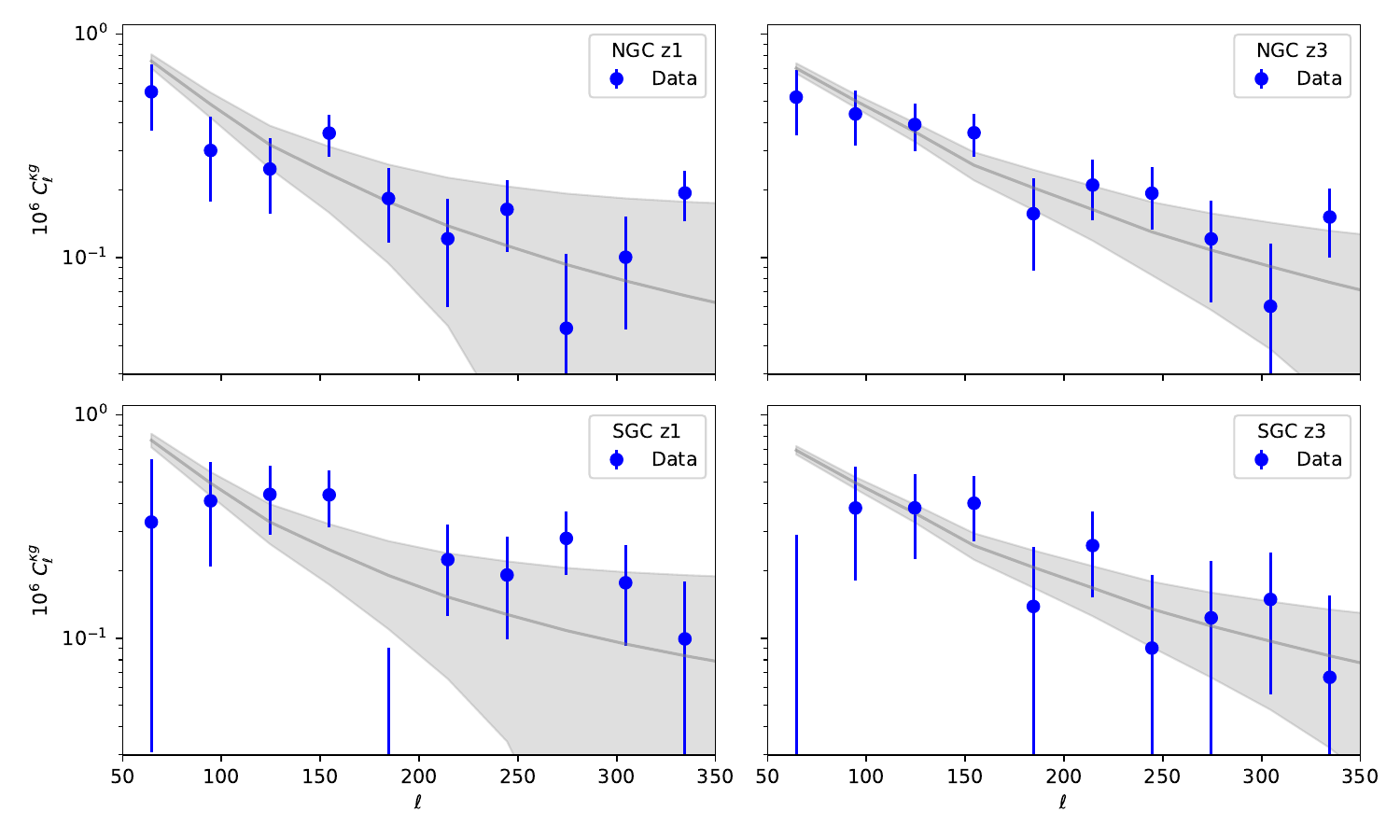}}
    \caption{The posterior-predictive distribution (grey bands) for the cross-correlation between BOSS galaxies and CMB lensing convergence, conditioned on the redshift-space galaxy clustering (including redshift-space distortions and baryon acoustic oscillations).  The measured cross-correlation (blue points) at high $\ell$ don't give much constraint within the context of perturbative models due to the combination of their large errors and the marginalization over counterterms (an effect which is more significant at lower redshifts i.e.\ \textbf{z1}).  The effect of the lensing data is thus largely a downward pull due to the low $\ell$ points. }
    \label{fig:ppd_lensing}
\end{figure}

A particular advantage of perturbative models of large-scale structure is that they rely on only a minimal set of theoretical assumptions to consistently model a wide range of clustering data. For example, the same bias parameters used to model the redshift-space clustering of BOSS galaxies in ref.~\cite{Chen22} also make robust predictions for their cross-correlation in weak lensing. Figure~\ref{fig:ppd_lensing} shows the posterior predictive distribution for these cross correlations, summarized as the angular multipoles of their 2-point function ($C^{\kappa g}_\ell$), with clustering and cosmological parameters conditioned on the redshift-space clustering data; those data tightly predict $C^{\kappa g}_\ell$ on large scales, while nonlinear bias and an additional effective-theory contribution to the matter-galaxy cross spectrum not probed by redshift-space clustering broaden the range of clustering amplitudes at smaller scales. Indeed, we can already see an intriguing feature of the joint BOSS and Planck data: the CMB lensing cross correlations at large scales (low $\ell$) are lower than what might be expected from the redshift-space clustering of BOSS galaxies, even after we marginalize over cosmology and nonlinear bias.  This is interesting, as the theoretical assumptions underlying the predictions are quite minimal: weak field gravity, at-most-weakly interacting and cold particle dark matter and a FLRW metric (by now well constrained by distance-redshift measurements). Figure~\ref{fig:ppd_lensing} also illustrates a more general feature of perturbative analyses of large-scale structure, which tend to extract cosmological information from large scales while\footnote{See e.g.\ Fig.~4 of ref.~\cite{Chen22} for a demonstration in the case of redshift-space clustering.} marginalizing over the transition to nonlinearity with bias and effective-theory parameters. Conversely, since additional information about these parameters cannot straightforwardly be gained by extending beyond the nonlinear scale, combining competing probes of the same structure (e.g. redshift-space clustering and weak lensing) can help better constrain these nuisance parameters by probing different combinations on perturbative scales.

The purpose of this paper is to demonstrate the viability of combined redshift-space and lensing analyses within perturbation theory using publically available data. In particular, we will use galaxies from the BOSS survey \cite{Dawson13} and CMB lensing maps from Planck \cite{PlanckLens18}, along with a theoretical model based on one loop (Lagrangian) perturbation theory \cite{Chen21}. We are not the first to look at this combination of data (see e.g.\ refs.\ \cite{Pullen16,Singh17,Doux18,Singh20}), but are, to our knowledge, the first to apply the full machinery of perturbation theory in this context, applying a consistent dynamical model without empirical prescriptions for galaxy clustering to model both the two and three-dimensional data, a technique which we expect will be critical given the significantly enhanced accuracy needs and scientific promise of the currently operating cosmological surveys like the Dark Energy Spectroscopic Instrument \cite{DESI}, the Atacama Cosmology Telescope \cite{Thornton16}, the South Pole Telescope \cite{Benson14} and their even more powerful successors. Since our main purpose is to perform a proof-of-principle study on public data, throughout this work we follow the BOSS collaboration's choices for systematics weights, masks and redshift binning in order to leverage the considerable effort that has gone into measuring the statistics, performing systematics checks and creating mock catalogs for covariance matrices for these samples, with only a few small, theoretically-motivated tweaks which we believe will be useful in future analyses.

The outline of the paper is as follows.  In the next section we discuss the data sets that we use.  Section \ref{sec:mocks} describes the mock catalogs used to validate our analysis pipeline, while section \ref{sec:theory} describes our theoretical models and assumptions.  Our results are presented in section \ref{sec:results}, along with a comparison to previous results.  We conclude in section \ref{sec:conclusions}, while some technical details of how we handle massive neutrinos are relegated to an appendix.

\section{Data}
\label{sec:data}

\subsection{BOSS Galaxies}

The BOSS survey \cite{Dawson13} is a spectroscopic galaxy survey part of the Sloan Digital Sky Survey III \cite{SDSSIII}, covering 1,198,006 galaxies over 10,252 square degrees of sky. Our analysis of the three-dimensional clustering of these galaxies follows that of \cite{Chen22}, which is described in detail in Section 2 of that work. Briefly, we follow the convention in ref.~\cite{Alam17} and split the BOSS galaxies into four independent samples, defining two redshift bins $0.2 < z < 0.5$  (\textbf{z1}) and $0.5 < z < 0.75$  (\textbf{z3}) split between the Northern (\textbf{NGC}) and Southern (\textbf{SGC}) galactic caps. In particular, we make use of the publicly available power spectrum, window function and mock measurements of each of these samples presented in ref.~\cite{Beutler21}. In order to better utilize the cosmological distance information in galaxy clustering, particularly through baryon acoustic oscillations (BAO), we will also use the post-reconstruction correlation functions measured in ref.~\cite{Vargas18}. Unlike the power spectra, these correlation functions were measured assuming that the $\textbf{NGC}$ and $\textbf{SGC}$ samples could be combined into one homogeneous sample. In order to take into account the cross-correlations between the power spectra and correlation function measurements, we construct our covariance matrix using measurements of these quantities in the V6C BigMultiDark Patchy mocks described in \cite{Kitaura16}; these measurements are also publicly provided by refs.~\cite{Beutler21,Vargas18}. Both power spectra and and correlation functions were computed assuming a fiducial cosmology with $\Omega_{m,\rm fid} = 0.31$.

In order to cross-correlate the BOSS galaxy density with CMB lensing, as described below, we also generate projected two-dimensional sky maps of the galaxy density. These maps are generated in the standard manner.  We first cut the galaxies to the desired hemisphere and redshift range (using the spectroscopic redshift).  Each galaxy is assigned a weight, $w_{\rm sys-tot}(w_{\rm cp}+w_{\rm no-z}-1)$, as described in detail in the BOSS papers \cite{Reid16,Ross17}.  The weighted counts of galaxies are computed in Healpix \cite{Gorski05} pixels at $N_{\rm side}=2048$ to form a ``galaxy map'' in galactic coordinates.  The random points supplied by the BOSS team are also binned into Healpix pixels to form the ``random map''.  The overdensity field is defined as the ``galaxy map'' divided by the ``random map'', normalized to mean density and mean subtracted.  We obtain a (binary) mask for the galaxies by keeping only those pixels where the random counts exceed 20\% of the mean random count (computed over the non-empty pixels) and overdensities outside of the mask are set to zero. We use the magnification bias slopes measured in \cite{Kramsta21}, viz.\ $s_{z1} = 0.77 \pm 0.02$ and $s_{z3} = 1.05 \pm 0.11$. 

Since the 2D (auto) clustering information within the galaxy map is a subset of that included in the 3D clustering measurements described above, we do not use the 2D galaxy angular power spectrum $C^{gg}_\ell$ derived from these maps except when estimating covariances, as described below. Not including $C^{gg}_\ell$ somewhat immunizes our analysis against purely angular systematics in the galaxy maps since, unlike $C^{gg}_\ell$ which only depends on line-of-sight angles $\mu \approx 0$, the clustering information in redshift-space multipoles are weighted across all $\mu$, though we note that our model fits either measurement in the data consistently.

\begin{figure}[htb]
    \centering
    \resizebox{0.92\columnwidth}{!}{\includegraphics{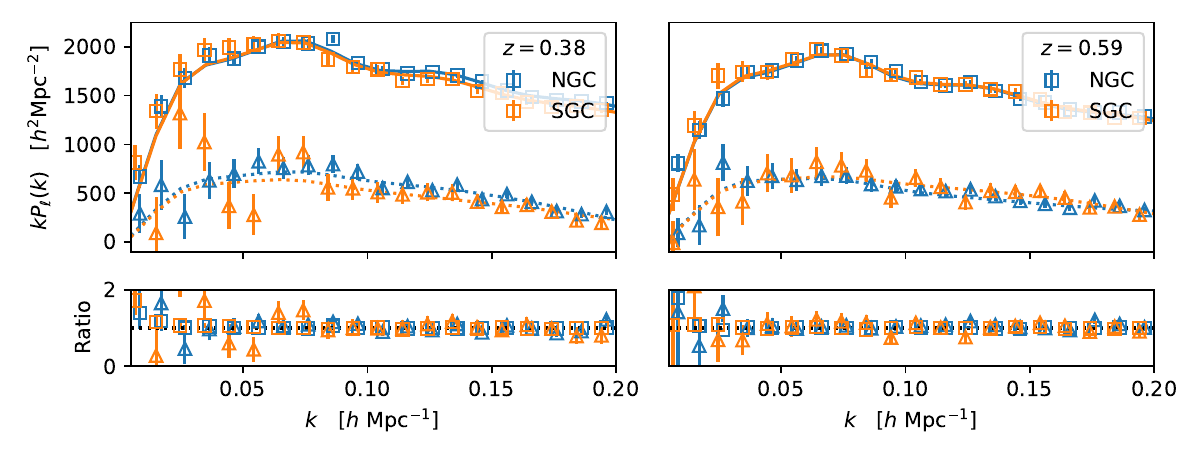}}
    \resizebox{0.92\columnwidth}{!}{\includegraphics{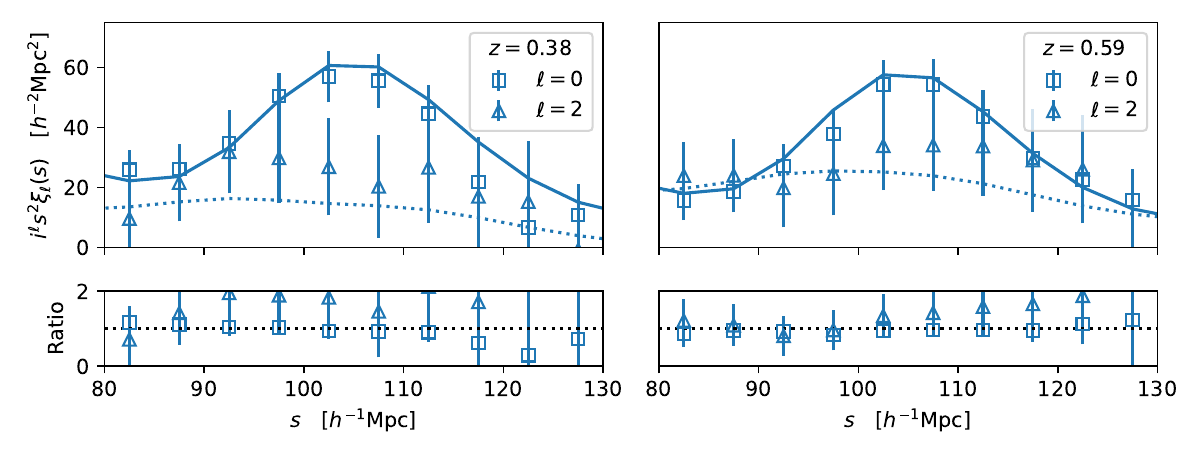}}
    \resizebox{0.92\columnwidth}{!}{\includegraphics{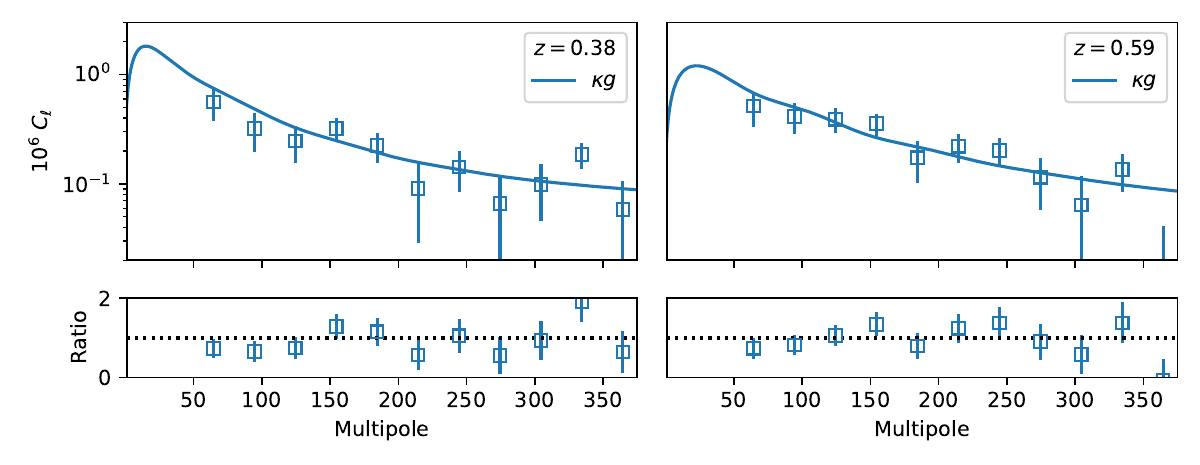}}
    \caption{The data to which we fit, in the form of 2-point functions vs.\ linear or angular scale.  The top row shows the pre-reconstruction redshift-space galaxy power spectrum multipoles for the two galactic hemispheres (NGC and SGC) and two redshift slices.  The middle row shows the post-reconstruction galaxy correlation function multipoles.  The bottom row shows the angular cross-spectrum between the galaxy overdensity and CMB convergence.  In each row the upper panels show the data while the lower panels show the ratio of the data to the best-fitting theoretical model (shown as the lines in the upper panels in each case). }
\label{fig:spectra}
\end{figure}

\subsection{Planck CMB Lensing}

Our treatment of the Planck CMB lensing maps is quite standard, and in detail follows that in refs.~\cite{Kitanidis21,White22}.  Specifically we use the 2018 Planck release \cite{PlanckLens18} available from the Planck Legacy Archive.\footnote{PLA: \url{https://pla.esac.esa.int/}}
These data are provided as spherical harmonic coefficients of the convergence, $\kappa_{\ell m}$, in HEALPix format \citep{Gorski05} and with $\ell_{\rm max} = 4096$.  We use the minimum-variance (MV) estimate obtained from both temperature and polarization, based on the \texttt{SMICA} foreground-reduced CMB map.  The maps are low-pass filtered and apodized as in ref.~\cite{White22} to produce a $\kappa$ map in HEALPix format at $N_{\rm side}=2048$.

Since the MV reconstruction in Planck is dominated by temperature, residual galactic and extragalactic foregrounds may contaminate the signal. Extensive testing has been performed by the Planck team, indicating no significant problems at the current statistical level \cite{PlanckLens18}.  However, as a test, we repeated the analysis with a lensing reconstruction provided by the Planck team that is based upon \texttt{SMICA} foreground-reduced maps where the thermal tSZ effect \cite{SZ72} has been explicitly deprojected \cite{PlanckLens18}. While they mitigate the effect of tSZ, these maps also tend to enhance the effect of other foregrounds like the cosmic infrared background (CIB) \cite{Sailer:2021vpm}. Swapping in these maps for the fiducial ones can therefore serve as a sanity check to test our sensitivity to residual foregrounds. We found that our results are very consistent between analyses, with the deprojected map leading to larger uncertainties, as expected.  This is in line with the expectation that extragalactic foregrounds lead to very small biases compared to our error bars \cite{vanE14,Osborne14,Baxter19,Sailer:2021vpm,Darwish21}, though those biases would typically be to lower $\kappa$ if the foregrounds have significant small-scale power (since they ``appear'' like a demagnified region).  We shall return to this in \S\ref{sec:results}. Additionally, the Planck lensing maps mask regions with SZ clusters, removing high-density regions; biases due to this effect are known to be very subdominant, however \cite{PlanckLens18}.

In order to estimate the cross-correlation of the CMB $\kappa$ map with BOSS galaxies, we use the pseudo-$C_\ell$ method \cite{Hivon02} as implemented within the \texttt{NaMaster} package \cite{Alonso18} to estimate our angular power spectra.  This technique is now very standard and has been described in detail elsewhere (e.g.\ refs.~\cite{Garcia21,White22} and the many references therein).  Briefly, this approach first computes the (pseudo) angular power spectrum as an average over $m$-modes of the spherical harmonic transform of the masked field.  The pseudo-spectra are binned into a discrete set of bandpower bins, $L$, and the mode coupling is deconvolved \cite{Hivon02}.  We use the \texttt{compute\_full\_master} method in \texttt{NaMaster} \cite{Alonso18} to calculate the binned power spectra and the bandpower window functions relating them to the underlying theory: $\langle C_L\rangle = \sum_\ell W_{L\ell} C_\ell$.
We choose a conservative binning scheme with linearly spaced bins of size $\Delta\ell = 30$ starting from $\ell_{\rm min} = 50$.  The bin width is larger than expected correlations between modes induced by the survey masks, while being narrow enough to preserve the structure in our angular spectra. To avoid power leakage near the edge of the measured range we perform the computation to $\ell = 6000$, and simply discard the bins beyond some $\ell_{\rm max}$ \cite{Krolewski20}.

\subsection{Covariance}

Throughout we shall make use of a Gaussian likelihood function with fixed covariance.  The covariance matrix for the three dimensional clustering (both pre- and post-reconstruction) is computed from mock catalogs supplied by the BOSS collaboration.  The covariance matrix for the lensing-galaxy cross-correlation is computed using \texttt{NaMaster} taking into account the disconnected contributions which dominate in the regimes of interest.

We neglect the covariance between the three dimensional clustering measures and the lensing-galaxy cross-correlation. Since the lensing kernel is so broad, the lensing-galaxy cross-correlation probes modes with very low wavenumber $k_\parallel$ along the line of sight, while the three dimensional clustering measurements are dominated by $k_\parallel\sim k$, leading to little overlap in Fourier space \edit{\cite{Taylor22}}. In addition, the lensing signal is predominantly from matter clustering at higher redshifts than the range we probe in this work, and moreover are dominated by noise in the temperature maps used for the lensing reconstruction over most of the $\ell$ range that we fit to.

\section{Mock catalogs}
\label{sec:mocks}

In this section we describe the N-body-based mock catalogs that we used to validate our analysis pipeline and compare their clustering to the BOSS data.  Since the mock catalogs used for pipeline validation were not used as inputs to the analysis (e.g.\ as part of the theory model or covariance calculation) but rather simply as validation tools the requirements on those mocks can be quite relaxed.  

Our analysis was not conducted blindly, because the catalogs and clustering measurements have long been public and we had previously done cosmology fits to the BOSS data alone \cite{Chen22}.  However, we did validate a number of the analysis choices on mock catalogs prior to performing the cosmology fits and we did not modify those choices when we fit to the data itself.

\subsection{Mock BOSS Catalogs}

Our mock catalogs are constructed from the \texttt{Buzzard v2.0} simulations \citep{DeRose2019, DeRose2021} in order to approximately reproduce the \textbf{z3} bin from the data. We do not use the \textbf{z1} bin from the simulations as the redshift range of this bin overlaps the transition between two distinct $N$-body simulations that the \texttt{Buzzard} catalogs are constructed from. Galaxies are included in these simulations using the \texttt{Addgals} algorithm \cite{Wechsler2021, DeRose2021b}, which assigns galaxies with mock SEDs, shapes and sizes to particles in the $N$-body lightcones. The spectra are integrated over the desired bandpasses to obtain broadband apparent magnitudes. The simulations are ray-traced in order to compute weak-lensing deflections, shears and magnifications for each galaxy. In order to select a CMASS-like sample from our simulations, we apply the CMASS color selection to our simulated catalogs, with minor adjustments to the color cuts that are tuned in order to better reproduce the redshift distribution of the \textbf{z3} sample. The effective redshift of our mock sample is $z_{\rm eff}=0.575$ and a magnification coefficient of $\alpha_{\rm mag}=1.3$. All mock measurements used in this work are the mean of 7 quarter-sky simulations.

\subsection{Power spectrum multipoles}

We compute mock spectra in our simulations using an independent pipeline from that used in the data. We compute $P_{\ell}(k)$ using the algorithm described in \cite{Hand2017}, making use of FKP weights and assuming the same $k$ binning as used in the BOSS data. In order to account for the effect of the window function, integral constraint, and wide-angle effects on our redshift-space clustering measurements, we follow the formalism described in \cite{Beutler21}, implemented in an independent pipeline from that used on the data, and validated against the BOSS DR12 NGC \textbf{z3} window and wide-angle matrices used in this work.

\subsection{Lensing cross-correlations}

We compute $C_\ell^{\kappa g}$ and mode coupling matrices from our simulations using \texttt{NaMaster}, using the same $\ell$ binning and mask apodization as that used in the data. The CMB lensing convergence field is computed using the Born approximation. We also use the same weighting scheme as applied to the BOSS data, including both FKP and inverse lensing kernel weights as described in section~\ref{sec:eff_redshift}.

\subsection{Post-reconstruction correlation functions} 

Non-linear evolution broadens the BAO peak in the correlation function, weakening the inferred distance constraints \cite{ESW07,Crocce08}.  However much of the broadening comes from large scales that can be well modeled and measured by a galaxy redshift survey.  The displacements induced by these large-scale modes can be inferred from the data and their impacts `undone', in a process known as reconstruction \cite{ESSS07}.  This has become a standard feature of BAO analyses, and was used throughout the BOSS survey \cite{Vargas18}.  We apply the same procedure to the mock catalogs.  We use \texttt{recon\_code}\footnote{\url{https://github.com/martinjameswhite/recon_code}}, adopting the isotropic BAO (or `Rec-Iso') \cite{White15,PWC09,Seo16} convention as in ref.~\cite{Vargas18}.  To form the overdensity field, the galaxy and random catalogs are converted to Cartesian coordinates using the correct distance-redshift relation for the cosmology of our mocks and deposited to grids using cloud-in-cell interpolation to a grid with $N_g = 512$ points in each dimension. The over-density field is $\delta_g(\bx) = \rho_{\rm gal} (\bx) / \rho_{\rm rand} (\bx) - 1$, where regions with $\rho_{\rm rand} = 0$ are set to zero by default. This field is then smoothed by a Gaussian kernel given by $\exp \left [ -( x/R_f )^2 /2 \right ]$, with $R_f = 15 \Mpc$, giving a smoothed field $\tilde{\delta}(\bx)$.  The displacement field, $\hat{\bPsi}^{\rm rec}$, is the solution of
\begin{equation}
    \partial_i \hat{\Psi}^{\rm rec}_i + \beta \partial_i \left ( \hat{r}_i \hat{r}_j \hat{\Psi}^{\rm rec}_j \right ) = \frac{\tilde{\delta}}{b}.
\end{equation}
where we have used $b=1.8$ and $f=0.872$ since the values used by BOSS were not given in ref.~\cite{Vargas18}.  This equation is solved using a multigrid relaxation technique with a V-cycle based on damped Jacobi iteration.  Both randoms and galaxies are then shifted by $\hat{\bPsi}^{\rm rec}$, with their appropriate factors in the ``Rec-Iso'' scheme, and their locations are converted back to angular coordinates and redshifts.

The multipoles of the correlation function $\xi_\ell (\bs)$ are measured from the above catalogs using \texttt{corrfunc} \cite{2020MNRAS.491.3022S}.  Denoting by $D$, $S$ and $R$ the shifted data and randoms and the fiducial random catalogs, respectively, the Landy-Szalay estimator
\begin{equation}
    \xi^{\rm rec} (s,\mu) = \frac{DD (s,\mu) - 2DS(s,\mu) + SS(s,\mu)}{RR (s,\mu)},
\end{equation}
is used to estimate $\xi^{\rm rec} (s,\mu)$.  We adopt linearly spaced bins of width $\Delta s = 5\Mpc$ and 100 bins in $\mu \in [0,1)$ following ref.~\cite{Ross17}. Multipoles of the correlation function are constructed by integrating in the $\mu$ direction.

\section{Theory Model}
\label{sec:theory}

In this work we aim to obtain cosmological constraints combining the three-dimensional distribution of galaxies in redshift space and the distribution of dark matter that they trace, reflected in its contribution to CMB lensing. To this end we will use Lagrangian perturbation theory (LPT), which models the gravitational clustering underlying RSD, BAO and CMB lensing within a unified dynamical framework. In the following subsections we describe how to connect this clustering with observables, provide a brief summary of LPT and describe how we efficiently emulate its predictions using Taylor series for the purposes of MCMC.

\subsection{Cosmological Parameters, Neutrinos and Linear Theory}
\label{ssec:lin_cosmo}

\begin{table}
    \centering
    \input{Tables/cosmo_priors}
    \input{Tables/cfixed}
    \caption{Cosmological parameter priors and values for our analysis. Uniform distributions are denoted $U(x_{\rm min},x_{\rm max})$.
    }
\label{tab:cpriors}
\end{table}

Throughout this paper, we will assume a $\Lambda$CDM cosmology with uniform priors on $\Omega_m$, $H_0$ and $\ln(10^{10} A_s)$ as described in Table~\ref{tab:cpriors}, with all other parameters fixed to their Planck best-fit values and assuming a minimal neutrino mass scenario $M_\nu = 0.06$ eV, mirroring the setup in ref.~\cite{Chen22}. Given such a set of cosmological parameters, we use CLASS \cite{CLASS} to compute the linear-theory power spectrum as the input to our one-loop perturbation theory. We operate within the EdS approximation wherein higher-order corrections scale linearly with powers of the linear power spectrum amplitude \cite{Takahashi:2008, Fasiello+:2016, delaBella:2017, Fujita+:2020, Donath+:2020}. 

In order to account for the effect of massive neutrinos we use the now-standard approximation that galaxies trace the cold dark matter and baryon $cb$ field \cite{Castorina15}, i.e.\ $\delta_g = \delta_g[\delta_{cb}]$ , which was recently shown to be an excellent approximation well into the quasilinear regime \cite{Bayer21}. Within this approximation the redshift-space galaxy power spectrum can be computed simply by plugging $P_{\rm cb,lin}$ as the linear power spectrum into the perturbation theory formulae.

\begin{figure}
    \centering
    \includegraphics[width=\textwidth]{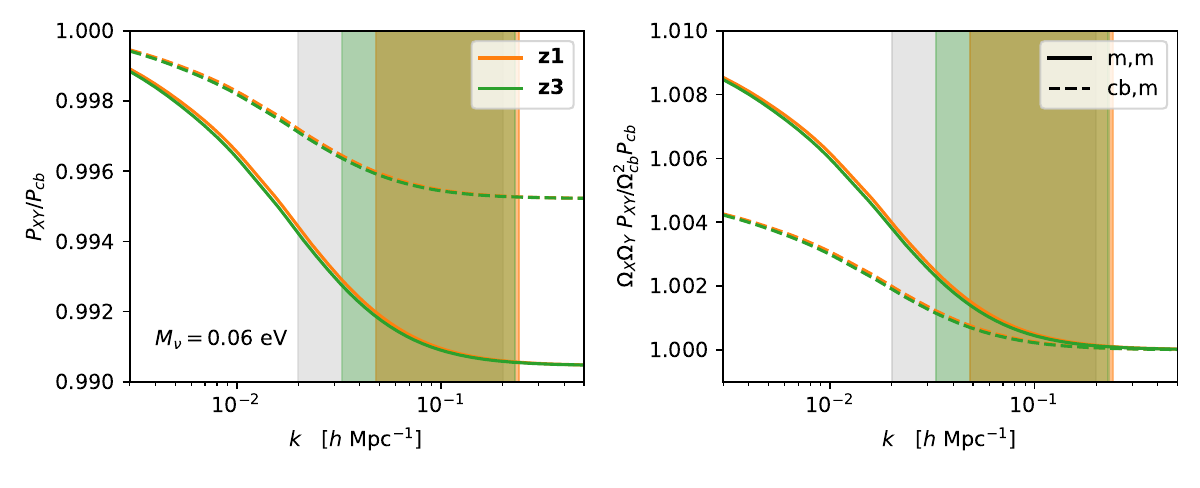}
    \caption{Ratios of linear theory total-matter power spectrum and matter and dark matter-baryon cross spectrum to the dark matter-baryon power spectrum (left), as well as the corresponding ratios for mass-weighted power spectra (right) in the Planck cosmology with minimal neutrino mass $M_\nu = 0.06$ eV. Green and orange shaded regions show the corresponding wave numbers probed by the lensing-galaxy cross correlations and grey regions show the wavenumbers probed in our RSD analysis. }
    \label{fig:nu_scales}
\end{figure}

Weak lensing-galaxy cross correlations, on the other hand, require a bit more care. In particular, as lensing is sourced by all matter, we must take the contribution from neutrinos into account explicitly. Indeed, this explicit neutrino mass dependence is precisely what allows galaxy-lensing cross correlations to be a potentially powerful probe of the neutrino mass \cite{Yu18,Bayer21}. At linear order this implies that $P_{gm} = b P_{cb,m}$, where $b$ is the Eulerian galaxy bias. Since neutrinos contribute negligible clustering power below the free-streaming scale, one approximation \cite{Krolewski20} is to use the fact that $\Omega_m \delta_m \approx \Omega_{cb} \delta_{cb}$ on these scales to make the substitution $\Omega_m P_{cb,m} \approx \Omega_{cb} P_{cb}$. However, as shown in Fig.~\ref{fig:nu_scales}, the quasilinear scales on which our analysis is based covers much of the transition region between the high $k$ modes with `unclustered' neutrinos and the low $k$ regime where neutrinos cluster with cold dark matter. In order to better capture this transition, we will instead compute perturbation predictions for the matter-galaxy cross power spectrum using $P_{cb,m}$ as the input linear power spectrum assuming the same bias coefficients as those in  $\delta_g = \delta_g[\delta_{cb}]$. As we show in Appendix~\ref{app:pt_nus}, such a scheme is accurate to order $f_\nu \mathcal{O}(\PL^2)$, that is of order the neutrino mass fraction times the typical (subleading) one-loop correction, and thus more than adequate for any upcoming analyses.

\subsection{Lagrangian Perturbation Theory}

Lagrangian perturbation theory models gravitational structure formation by following the displacements $\Psi(\bq,\tau)$ of fluid elements starting at Lagrangian positions $\bq$ at the initial time. These displacements follow Newtonian gravity in expanding spacetimes $\ddot{\Psi} +  \mathcal{H} \dot{\Psi} = - \nabla_{\bx} \Phi$, where dots are with respect to conformal time, and map the initial positions of fluid elements to their observed ones via $\bx = \bq + \Psi(\bq,\tau)$. The Newtonian potential $\Phi$ is in turn sourced by the overdensity $\delta_m$ of fluid elements under this evolution, given via mass conservation to be \cite{Mat08a}
\begin{equation}
    1 + \delta_m(\bx, \tau) = \dtq \delta_D(\bx - \bq - \Psi(\bq,\tau)).
\end{equation}
Within this framework the displacements are then solved order-by-order, that is perturbatively, in the initial conditions, i.e.\ $\Psi = \Psi^{(1)} + \Psi^{(2)} + \Psi^{(3)} + ...$, with the first-order solution commonly referred to as the Zeldovich approximation \cite{Zel70,Ber02}. Finally, within the effective-theory approach of LPT the effect of short-wavelength densities and velocities are integrated out, resulting in free counterterms and stochastic contributions whose form are restricted by symmetries but whose values must be fit to data and cannot be determined a priori \cite{PorSenZal14,VlaWhiAvi15}.

To model the distribution of galaxies we need to account for the fact that galaxies are imperfect and nonlinear tracers of matter. In the Lagrangian picture this is accomplished by writing the initial number density of proto-galaxies $F(\bq)$ as a local functional of the initial conditions. These proto-galaxies are then advected to their observed positions to give
\begin{equation}
    1 + \delta_g(\bx,\tau) =  \dtq F(\bq)\ \delta_D(\bx - \bq - \Psi(\bq,\tau)).
    \label{eqn:delta_g}
\end{equation}
In this paper we follow \cite{Chen22} and use the form \cite{Mat08b,CLPT,VlaCasWhi16,ChenVlahWhite20}
\begin{equation}
    F(\bq) = 1 + b_1 \delta_{m,0}(\bq) + \frac{1}{2} b_2 (\delta_{m,0}^2(\bq) - \avg{\delta_{m,0}^2}) + b_s (s_0^2(\bq) - \avg{s_0^2}).
    \label{eqn:lagbias}
\end{equation}
In particular we will operate under the assumption that third-order Lagrangian bias is small for small-to-intermediate mass halos \cite{Abidi18,Lazeyras18} and, along with the lowest-order derivative bias $\propto \nabla^2 \delta_{m,0}$, highly degenerate with counterterms. The matter density is equivalent to a tracer with all the Lagrangian bias parameters equal to zero, i.e.\ $F(\bq)=1$.

Our analysis in this paper specifically requires LPT predictions for the matter and galaxy two-point functions, in real and redshift space, pre- and post-reconstruction. Briefly, from Equation~\ref{eqn:delta_g} the power spectrum can be written as
\begin{equation}
    (2\pi)^3 \delta_D(\bk) + P(\bk) = \dtq \avg{e^{i\bk \cdot(\bq + \Delta)} F(\bq_1) F(\bq_2)}_{\bq = \bq_1 - \bq_2},
    \label{eqn:pktheory}
\end{equation}
where the pairwise displacement is given by $\Delta = \Psi(\bq_1) - \Psi(\bq_2)$.  To compute the power spectrum in redshift space, wherein line-of-sight distances are inferred from redshifts and hence include a contribution from peculiar velocities $\dot{\Psi}$, simply requires swapping in redshift-space displacements $\Psi_s$ boosted by line-of-sight velocities. 

A notable feature of Equation~\ref{eqn:pktheory} is that the exponentiation of the pairwise displacement allows for resummation of long-wavelength (IR) displacements which captures important physical effects such as the nonlinear damping of the BAO peak; to maintain consistency between the pre-reconstruction galaxy-galaxy and matter-galaxy power spectrum predictions we use the scheme proposed in ref.~\cite{Chen21} wherein linear displacements and velocities below $k_{\rm IR} = 0.2 \kMpc$ are resummed while shorter-wavelength modes are perturbatively expanded. This scheme is different than the one used in refs.~\cite{Kitanidis21,White22}, where all the linear displacements were resummed, but has been tested extensively in simulations and mocks \cite{Chen21,Chen22}\footnote{In particular, we use the $\mu=0$ output of the redshift-space power spectrum in \texttt{velocileptors} for the real-space power spectrum.}. In a similar vein, since reconstruction subtracts part of the large-scale displacements responsible for the nonlinear damping of the BAO, computing the two-point function after reconstruction requires the displacements in Equation~\ref{eqn:pktheory} have these subtracted as well; following \cite{Chen22} we will in addition make use of a saddle point approximation at the BAO scale to model the BAO damping form in the post-reconstruction correlation function, using a broadband model linear in $1/r$ to capture any residual smooth contributions. Our calculations throughout this work make use of the publically available code \cite{Chen20} \texttt{velocileptors}\footnote{https://github.com/sfschen/velocileptors}; we refer the interested reader to refs.~\cite{Chen21} and \cite{ChenVlahWhite19} for further discussions on modeling redshift-space distortions and reconstruction within LPT, respectively.

\subsection{Galaxy Clustering in 2D and 3D}

The observables we analyze in this work --- 3D clustering in redshift surveys and 2D angular cross-correlations with weak lensing from CMB experiments --- jointly probe matter and galaxy clustering within the cosmological volume surveyed by BOSS. While they reflect the same underlying clustering, however, the particularities of each measurement are sufficiently different that it is worth describing in some detail the connection between this clustering and each observable.

\subsubsection{Redshift-Space Clustering}

The 3D galaxy correlation function and power spectrum multipoles are measured in dimensionful coordinates ($\mathbf{r}$ and $\mathbf{k}$ respectively)--- a cosmological model must thus be assumed to convert angles and redshifts into comoving distances.  For BOSS the Cartesian coordinates of the galaxies were computed assuming a fiducial $\Lambda$CDM cosmology with $\Omega_{M,\rm fid}=0.31$.  This implies that when we test a model with a different redshift-distance relation to the fiducial model we must apply a rescaling of distances relative to the ``true'' cosmology in directions parallel and perpendicular to the line of sight.  This is often referred to as the Alcock-Paczynski effect \cite{Alcock79,Padmanabhan08} and is included in our model for the power spectrum and post-reconstruction correlation function.  Finally, the observed Fourier-space clustering power of galaxies in redshift surveys is the convolution of the true power with the survey window function; to take into account this and wide angle effects we adopt the formalism and data outputs of ref.~\cite{Beutler21}. Our treatment of these steps is identical to that in ref.~\cite{Chen22}, to which we refer readers seeking further details.

\subsubsection{Angular Power Spectra}

The 2D galaxy-lensing cross correlation, on the other hand, is reported in dimensionless angular coordinates. Within the Limber approximation \cite{Limber:1953} the angular multipoles are related to the matter-galaxy cross power spectrum $P_{mg}$ by
\begin{equation}
    C_\ell^{\kappa g} = \int d\chi\  \Bigg( \frac{W^{\kappa}(\chi) W^g(\chi)}{\chi^2} \Bigg)\ P_{mg}\Big( k=\frac{\ell + 1/2}{\chi},z \Big);
\end{equation}
All the dependence on cosmological distances is implicit in this integral such that the end result is independent of any fiducial cosmology. The galaxy and lensing kernels are given by
\begin{equation}
    W^g(\chi) = H(z) \frac{dN}{dz}, \quad W^{\kappa}(\chi) = \frac{3}{2} \Omega_m^2 H_0^2  (1+z) \frac{\chi (\chi_\ast - \chi)}{\chi_\ast},
\end{equation}
where $dN/dz$ is the weighted galaxy distribution and  $\chi_\ast$ is the distance to last scattering. In addition to the above term the projected galaxy density also receives a contribution from the so-called magnification bias. The magnification term is only a small contribution to the total signal, but because it probes the line of sight all the way to small radial distances it is sensitive to smaller scales than the other contributions.  We make use of the \texttt{HaloFit} fitting function for $P_{mm}$ (\cite{Smith03,Takahashi12} as implemented in \texttt{CLASS}) for the magnification contributions.

\subsubsection{Effective Redshift}
\label{sec:eff_redshift}

Both the two- and three-dimensional measurements above average galaxy and matter clustering over large spans of redshift (\textbf{z1} and \textbf{z3}) over which the universe expands by up to $20\%$, with comparable changes in other cosmological quantities like the linear growth factor, $D(z)$. In order to account for the evolution of both the background and galaxy sample we will make use of the effective-redshift approximation which we will now describe in some detail.

Defining the auto- and cross-spectra of each sample to evolve with redshift as $P_{gg,g\kappa}(\bk,z)$, both the two- and three-dimensional power spectra in this work can be written in the form
\begin{align}
    \hat{\Theta} &= \sum_i w_i P(\bk(z_i), z_i) \\
    &= \sum_i w_i \Big( P(\bk_i, z_{\rm eff}) + (z_i - z_{\rm eff}) \partial_z P(\bk_i, z_{eff}) + \half (z_i - z_{\rm eff})^2 \partial^2_z P(\bk_i, z_{\rm eff}) + ... \Big)
    \label{eqn:taylor}
\end{align}
The effective redshift\footnote{We will follow convention and use the redshift, $z$, as the `time' coordinate. This is not a unique choice and it is in principle possible to adopt other time coordinates, for example the scale factor $a$ --- the varying accuracy of the ``effective time'' approximation, i.e.\ dropping all but the leading term in Equation~\ref{eqn:taylor}, then depends on the size of the quadratic correction. As an example, in linear theory where we have that the matter power spectrum scales approximately as $D^2(z) \sim a^2 \sim (1 + z)^{-2}$, the error incurred by adopting $z_{\rm eff}$ would be three times larger than if we had instead chosen $a_{\rm eff}$. Since $b(z) D(z)$ and $f(z) D(z)$ are both extremely flat functions of redshift, we expect our analysis to be insensitive to this choice, though we do note that in the same limit $C^{\kappa g} \propto b(z) D^2(z) \sim a$ is linear in the scale factor, suggesting that using $a_{\rm eff}$ might be somewhat better for future analyses with greater accuracy needs.}, $z_{\rm eff}$, is then defined such that the linear term cancels, i.e.\ $z_{\rm eff} = \sum_i w_i z_i$. For example, cross-correlation of galaxies and CMB $\kappa$ has \cite{Modi17b}
\begin{equation}
    z^{\rm xcorr}_{\rm eff} = \int \frac{d\chi}{\chi^2}\ W^{g}(\chi) W^{\kappa}(\chi) z(\chi).
\end{equation}
Similarly the galaxy auto-spectrum has an effective redshift given by \cite{Matarrese97,White08,White12,Zhu15,deMattia21}
\begin{equation}
    z^{\rm 3D}_{\rm eff} = \frac{\int d^3 \br\ \bar{n}(\br)^2 z(\br)}{\int d^3 \br\ \bar{n}(\br)^2 } = \frac{ \sum_i w_i \bar{n}_i z_i  }{ \sum_i w_i \bar{n}_i  }
\end{equation}
where the sum is over the galaxies, $\bar{n}$ is the galaxy number density accounting for systematic and FKP weights, and $w_i$ are the product of the weights on each galaxy. The second equality above uses that $\int d^3\br\ \bar{n} = \sum_i w_i$. The above definition is distinct from, and more accurate when used on two-point clustering statistics \cite{Pryer21}, than the one defined in the official BOSS analysis \cite{Beutler17}.  The BOSS analyses used the mean redshift, written as
\begin{equation}
    z_{\rm mean} = \frac{\sum_{ij}w_iw_j (z_i+z_j)/2}{\sum_{ij} w_iw_j}
    = \frac{1}{2}\frac{\sum_i w_iz_i\sum_j w_j}{\sum_i w_i\sum_j w_j} + \frac{1}{2}\frac{\sum_j w_jz_j\sum_i w_i}{\sum_j w_j\sum_i w_i}
    = \frac{\sum_i w_iz_i}{\sum_i w_i}
\end{equation}
Since the product $fD$ is quite slowly varying, the difference in $z_{\rm eff}$ and $z_{\rm mean}$ is of little import for analyses of redshift-space distortions or baryon acoustic oscillations.  However it is potentially more important for measurements depending upon $D(z)$ itself, such as our lensing-galaxy cross-correlation.  Comparing these definitions, the linear growth factor $D(z)$ is $1\%$ higher at $z_{\rm eff} = 0.59$ compared to $z_{\rm mean}=0.61$ for the $\textbf{z3}$ bin, though they agree to within a fifth of a percent for $\textbf{z1}$ ($z_{\rm eff} \simeq z_{\rm mean} \simeq 0.38$).

The above discussion makes clear that the two and three-dimensional clustering analyzed in this work primarily reflect galaxy clustering at $z^{\rm xcorr}_{\rm eff}$ and $z^{\rm 3D}_{\rm eff}$, respectively. These can in principle be quite different; for the fiducial BOSS samples they are $0.367$ and $0.380$ for $\textbf{z1}$ and $0.589$ and $0.602$ for $\textbf{z3}$, respectively. Since we are interested in using the shared galaxy clustering from the two statistics, we take the additional step to weight the galaxies in the lensing cross correlation such that $z^{\rm xcorr}_{\rm zeff} \approx z^{\rm 3D}_{\rm zeff}$. In particular, we weight each galaxy by an additional factor $w(z) = W^{g}/W^\kappa$ (calculated assuming $\Omega_m^{\rm fid} = 0.31$) such that $W^g W^\kappa \rightarrow (W^g)^2$ and the cross-correlation has the same effective redshift as the (un-weighted) auto-correlation. To maintain consistency between the 2D and 3D clustering statistics it is important to employ the same set of weights in each; in particular, each galaxy receives the same set of systematic and FKP weights when computing the power spectrum, correlation function and angular multipoles. The two- and three-dimensional autocorrelation effective redshifts, as defined above, are equivalent. For example, for the 2D autocorrelation we have
\begin{equation*}
    z^{\rm 2D}_{\rm eff} = \frac{1}{\mathcal{N}} \int \frac{d\chi}{\chi^2}\ \left[W^{g}(\chi)\right]^2\, z(\chi)
    = \frac{1}{\mathcal{N}} \int \frac{d\chi}{\chi^2}\ \left[ H(z) \frac{dN}{dz} \right]^2\, z(\chi)
    \quad\mathrm{with}\quad
    \mathcal{N} = \int \frac{d\chi}{\chi^2} \left[W^{g}\right]^2 .
\end{equation*}
Since $H\, dN/dz \propto \chi^2\, \bar{n}$ and $d^3\br \propto \chi^2 d\chi$ the integral in the numerator reduces to $\int d^3\br\ \bar{n}^2\, z$, i.e.\ given the same galaxy weights and distributions, $z^{\rm 2D}_{\rm eff} =  z^{\rm 3D}_{\rm eff}$. We have checked that our weighting leads to effective redshifts agreeing to within a tenth of a percent for the cosmologies of interest in this work. 

\subsection{Gravitational slip}
\label{sec:slip}

Within general relativity, weak lensing and redshift-space distortions jointly probe the amplitude of matter clustering through gravity's effect on the trajectories of massless (photons) and massive (galaxies) particles. In principle, photon and galaxy trajectories are influenced by different components of the metric, the latter by the Newtonian potential $\Psi$ and the former by the Weyl potential $(\Phi+\Psi)/2$ --- these are equal at late times within General Relativity but could be different in modified theories of gravity. To test for such differences we include a free factor multiplying the amplitude of the lensing-galaxy cross correlation,
\begin{equation}
    c_\kappa = \frac{1 + \gamma}{2} \quad ,
\end{equation}
where $\gamma=\Phi/\Psi$ is the gravitational slip (see ref.~\cite{Joyce16} and references therein), and similarly the magnification bias-CMB lensing cross correlation by $c_\kappa^2$. Since our analysis is sensitive to the relative amplitude difference between redshift-space clustering and lensing cross-correlations, any deviation of the fit $\gamma$ from unity could indicate departures from general relativity in either velocities or gravitational lensing. Were we to free the neutrino mass within our analysis, this effect would be somewhat degenerate with the additional suppression of matter clustering due to free-streaming neutrinos --- our constraint on $\gamma$ therefore will also serve as some indication of our ability to constrain the neutrino mass through combining galaxy clustering with CMB lensing.\edit{\footnote{In particular, if we think of the lensing amplitude as probing $\delta_{m} \approx (1 - f_\nu) \delta_{cb}$ and the RSD as probing $f(z)\ \sigma_{8,cb}(z) \sim f_{M_\nu=0}(z)\ (1 - \frac{3}{5} f_\nu)\ \sigma_{8,cb}(z)$ then the relative amplitude of the lensing to RSD compared to the case where $M_\nu=0$ is $c_\kappa \sim 1 - \frac{2}{5} f_\nu$.}} Alternatively, within a fixed physical model comparing the relative amplitudes of the lensing and RSD signals through $c_\kappa$ allows us to perform a consistency check between the two datasets and check for systematics, akin to the scaling parameter $X_{\rm lens}$ multiplying cross spectra in the DES Y3 $3\times 2$pt analysis \cite{DESY3}.

\subsection{Emulators}

We use the now-standard method of Markov Chain Monte Carlo to explore the posterior distribution of our parameters.  In a high dimensional parameter space such as ours this involves many likelihood evaluations.  In order to minimize the computing resources we require, we replace the model calculations involved in the likelihood computation with an emulator based on Taylor series expansions \cite{DAmico20,Colas:2019ret,Mergulhao21,Chen22,DeRose22}. This reduces the time-per-likelihood-evaluation to tens of milliseconds.  Evaluation of the Taylor series coefficients is very fast (a few minutes per spectrum on one node of the \texttt{Cori} machine at NERSC\footnote{\url{www.nersc.gov}}).  Using a $4^{\rm th}$ order Taylor series, with coefficients computed by finite difference from a $9^3$ element grid centered around $\Omega_m=0.31$, $h=0.68$ and $\log(10^{10}A_s)=2.84$ we achieve an accuracy of better than $10^{-3}$ for real-space power spectra and the redshift-space monopole, and $10^{-2}$ for the redshift-space quadrupole, or better than $10^{-3}$ at $k < 0.1 \kMpc$, in terms of 68$^{\rm th}$ percentile fractional residuals, corresponding to less than one-tenth of the statistical error in any entry in our data vector. To further speed up model evaluations we emulate the (un-windowed) two and three-dimensional clustering directly --- given a set of cosmological parameters we predict the bias contributions to $P_\ell$, $\xi_\ell$ and $C^{\kappa g}$ taking into account the effective redshifts, fiducial distances and redshift kernels assumed for each sample.

\section{Results}
\label{sec:results}

Having laid out both the theory models and data measurements in the previous sections we are now in a position to extract cosmological constraints from the combined BOSS and Planck data. Since, unlike in the case of pure spectroscopic data, our methodology has not been previously tested, and in the view of preparing for the next-generation of cross-correlations analyses, we will proceed cautiously, starting by validating our theory model against the mock data described in \S\ref{sec:mocks} and performing sanity checks on the data before describing the cosmological constraints themselves.

\subsection{Priors and Scale Cuts}
\label{ssec:priors}

We begin by defining the scales over which we will fit the data. For the BAO and RSD data we largely follow ref.~\cite{Chen22}, fitting the pre-reconstruction monopole and quadrupole moments of the power spectrum for $0.02<k<0.20\,h\,\mathrm{Mpc}^{-1}$ and the post-reconstruction monopole and quadrupole moments of the correlation function for $80<s<130\,h^{-1}\mathrm{Mpc}$.  These scale cuts have been extensively validated against simulations to show that our perturbative model works to the desired accuracy within them. For the angular cross-clustering, $C_\ell^{\kappa g}$, we choose $\ell_{\rm max}=250$ for the \textbf{z1} slice and 350 for the \textbf{z3} slice.  We have chosen these conservatively to correspond to $k_{\rm max}\approx 0.20\,h\,\mathrm{Mpc}^{-1}$, the same scale cut we use for the redshift-space analysis, at the distance implied by the mean redshift of each sample. Our results are not very sensitive to either choice because the Planck $\kappa$ maps are very noisy at these scales.

In addition to the cosmological parameters (Table \ref{tab:cpriors}), our model contains numerous bias parameters, counter terms and stochastic terms for each redshift slice and galactic cap.  The priors we adopt for these are given in Table \ref{tab:priors}, and are based on those adopted in ref.\ \cite{Chen22} with two exceptions: we have narrowed the counterterm $\alpha_n$ priors in the view that they are in any case sufficiently well constrained by the data that the priors are uninformative, and that they should represent only modest corrections to linear theory on scales where perturbation theory is valid. We have also updated the prior on the isotropic stochastic term $R_h^3$ for the higher-redshift sample to better reflect the effective number density of the $\textbf{z3}$, where $\bar{n}^{-1} \approx 6000 \Mpccube$, such that the priors on $R_h^3$ in both \textbf{z1} and \textbf{z3} reflect the latest studies on stochasticity in BOSS-like galaxies \cite{Kokron21}. Adopting these new priors shift our constraints on $\sigma_8$ by roughly $0.2\,\sigma$, with all other parameters essentially unaffected, compared to ref.~\cite{Chen22}.

\begin{table}
    \centering
    \input{Tables/priors2}
    \input{Tables/priors}
    \caption{Perturbation theory (left) and nuisance broadband parameter (right) priors and values for our analysis. Uniform distributions are denoted $\mathcal{U}(\mathrm{min},\mathrm{max})$ while normal distributions are denoted by $\mathcal{N}(\mu,\sigma)$. The prior isotropic stochastic term $R_h^3$ has its width set to one-third of Poisson value after shot-noise subtraction --- this is roughly $\bar{n}^{-1} \approx 3000\, h^{-3}\mathrm{Mpc}^3$ for $\textbf{z1}$ and $6000\, h^{-3}\mathrm{Mpc}^3$ for $\textbf{z3}$.
    }
\label{tab:priors}
\end{table}

\subsection{Tests on Mock Data}
\label{ssec:mock_results}

\begin{figure}
    \begin{minipage}{.45\textwidth}
    \centering
    \includegraphics[width=\textwidth]{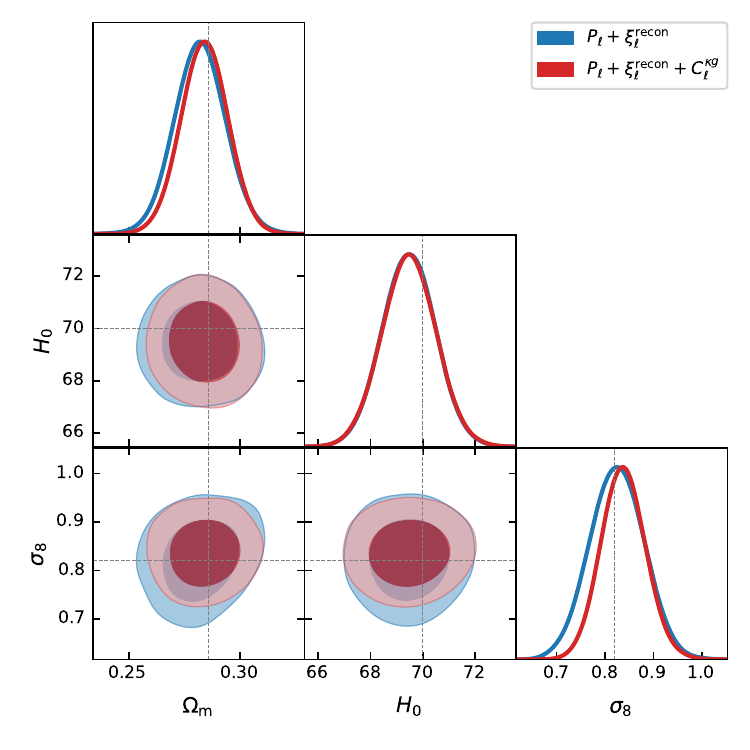}
    \end{minipage}\hfill
    \begin{minipage}{.52\textwidth}
    \centering 
    \input{Tables/mock_results}
    \end{minipage}
    
    \caption{(Left) Mock constraints from the mean of the Buzzard mocks for the \textbf{z3} sample fitting redshift-space power spectrum and post-reconstruction correlation function, with and without (red and blue) galaxy-CMB lensing cross-correlations multipoles. Gray lines indicate the true cosmology of mocks given by $\Omega_m = 0.286$, $h = 0.7$ and $\sigma_8 = 0.82$. (Right) Summary of the mock constraints (mean$\pm 1\,\sigma$). Adding angular cross-correlations to the data vector improves $\sigma_8$ constraints by close to $20\%$.  }
    \label{fig:mock_results}
\end{figure}

While the models we use in this paper have been tested extensively on mocks and data in the context of both spectroscopic surveys \cite{Chen21,Chen22} and angular cross correlations of galaxy clustering and lensing \cite{Modi17b,White22}, they have not been tested on the combination of these data as required for this work. In this subsection we use the mock data described in \S\ref{sec:mocks} to test whether LPT can indeed jointly and consistently model the matter and galaxy clustering encoded in our data to the required accuracy. To this end we apply the same pipeline, swapping only the input data vectors, that we will apply to the observed data, with the same scale cuts and priors. By necessity, this test only covers one redshift bin (\textbf{z3}), and the fixed cosmological parameters ($\omega_b, n_s, M_\nu$) have been adjusted to those of the mocks.

Our results are shown in Figure~\ref{fig:mock_results}. Fits using LPT recover the true cosmology of the Buzzard mocks to well within $1\,\sigma$ both before (blue, left) and after (red, right) the addition of angular galaxy-lensing cross correlations. Indeed, the implied means of both $\Omega_m$ and $\sigma_8$ fall within $0.38\,\sigma$ of the truth, roughly the expected statistical deviation for these mocks given that the Buzzard mocks cover 7 times the sky area of BOSS\footnote{We note that this scaling is not exact for a number of reasons, including that the redshift-space quadrupole is roughly $25\%$ larger than that in the BOSS data, and therefore noisier than the covariance matrix we use might imply, and that the lensing maps in the simulations are in principle ``noiseless'' and therefore cosmic-variance dominated at all scales, unlike the Planck maps for which this is true only at large scales ($\ell \lesssim 150$) --- which are however the scales from which most of the $\sigma_8$ constraint is derived.}. The Hubble parameter $H_0$ falls $0.5\,\sigma$ from truth, also not inconsistent with statistical scatter, especially since the $H_0$ constraint derives almost entirely from redshift space and previous tests on simulations \cite{Chen21,Chen22} with far lower statistical scatter have shown that our model can recover unbiased $H_0$ in these cases. These results therefore validate our perturbation theory modeling of the underlying gravitational nonlinearities studied in this work. In addition, including angular CMB-lensing and galaxy cross correlations improves the $\sigma_8$ constraint from these mocks by close to $20\%$, and the $\Omega_m$ constraint by $10\%$ --- even given the relatively noisy Planck lensing data --- demonstrating the potential gains from cross-correlations analyses like ours.

\subsection{Systematics Checks and Analysis Setup}
\label{sec:systematics}

\begin{figure}
    \centering
    \includegraphics[width=\textwidth]{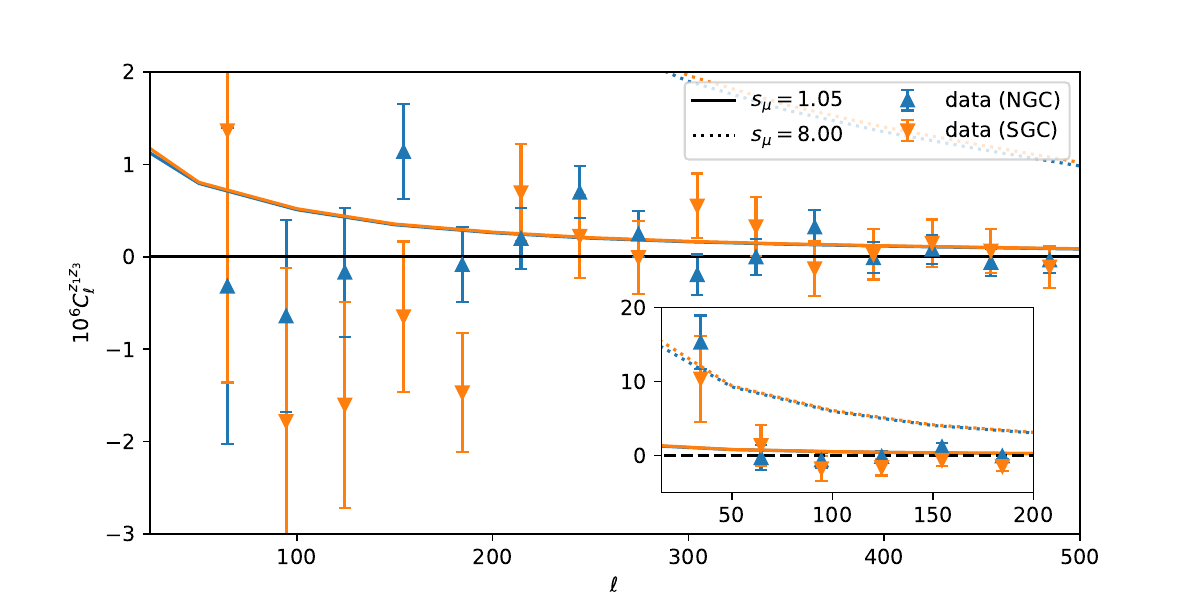}
    \caption{The (angular) cross-power-spectra for the low (\textbf{z1}) and high (\textbf{z3}) redshift galaxy samples.  Since the two maps are disjoint in redshift, any signal should be dominated by magnification of the higher redshift sample (solid lines).  The main panel shows the cross-correlation in the NGC (blue) and SGC (orange) respectively, while the inset shows a zoom-out on the $y$-axis to capture the large signal seen at $\ell<50$.  The dotted lines show the size of the magnification bias predicted for $s_\mu\approx 8$.  }
    \label{fig:mag_bias_check}
\end{figure}

To check for non-cosmological contributions to the projected clustering in the BOSS galaxy maps, we cross-correlate the \textbf{z1} and \textbf{z3} samples in both the NGC and SGC.  Since these maps are separated in redshift, with galaxy redshifts determined spectroscopically, in the absence of systematics the cross-correlation signal should be dominated by the effects of magnification. A similar test was conducted in ref.~\cite{Doux18}, who cross correlate the LOWZ and CMASS samples (which were combined to form jointly form the \textbf{z1} and \textbf{z3} samples used in this work) and find no significant evidence of correlations due to either systematics or magnification bias.  This is not true for the \textbf{z1} and \textbf{z3} samples, as we show in Fig.~\ref{fig:mag_bias_check}. This apparent discrepancy could potentially be due to the fact that the BOSS systematic weights were computed to normalize the angular distributions of LOWZ and CMASS individually and account for effects like stellar density and seeing --- however this re-weighting may not be optimal for the combined sample, split by redshift, if the weights are not readjusted for this purpose, as we show in Appendix~\ref{app:wsys_z}, particularly if the effects of the systematics are redshift-dependent. Any such angular systematics can in turn correlate with the CMB $\kappa$ map and bias our results.

Concentrating first on the inset panels of Fig.~\ref{fig:mag_bias_check} we see a very large cross-correlation at $\ell<50$, that is inconsistent with the expected size of any magnification signal.  In order to match the amplitude seen at $\ell<50$ the slope of the number counts in the \textbf{z3} slice would need to be $s_\mu\approx 8$, which would then result in a signal grossly inconsistent with the points at $\ell>50$.  Given the rapid drop in cross-power with $\ell$ we suspect this contamination may be galactic in origin.  To isolate ourselves from this effect, we have chosen $\ell_{\rm min}=50$ when computing $C_\ell^{\kappa g}$ (\S\ref{sec:data}).

The second thing to note in Fig.~\ref{fig:mag_bias_check} is the negative cross-correlation for $50<\ell<200$ in the SGC, with no such signal in the NGC.  Such an anti-correlation is unlikely to arise from magnification given reasonable slopes for the number counts, $s_\mu$.  The signal is well detected, statistically, and covers the whole range of scales where we expect significant $S/N$ in our cross-correlation signal.  We do not know the cause of this anti-correlation, and we are not certain that this systematic would correlate with the CMB lensing signal.  Out of an abundance of caution, and because the SGC contains relatively little statistical weight overall, we choose to drop the lensing cross-correlation in the SGC from our data vector, retaining only the NGC data.

It is worth noting\footnote{We thank Ashley Ross for pointing out this potential source of systematics.} that the LOWZ sample comprising most of the galaxies in \textbf{z1} contains early ``chunks'' in the NGC selected using a slightly different algorithm than later ones \cite{Reid16,Ross17}. These data could potentially have different systematics than the later chunks, and indeed one of them was found to require corrections based on seeing. Simply masking these data in the cross-correlation is not possible however, as they may represent a different subset of galaxies than the full sample and our assumption that a single set of biases describes both the redshift-space and projected clustering would be invalidated. In order to make use of the publicly available clustering data released by the BOSS collaboration, including window functions and mocks, thus depends critically on the collaboration's determination that the combined sample is sufficiently uniform after the corrections they performed \cite{Reid16,Ross17}.

As an additional check\footnote{We thank Anton Baleato Lizancos for suggesting this test.} we cross-correlated a map constructed from the systematics weights applied to the galaxies with the Planck $\kappa$ map.  In principle there should be no correlation, but we find something small but non-zero for both \textbf{z1} and \textbf{z3}.  This must arise due to correlations between signal, foregrounds or noise patterns in the $\kappa$ map that correlate with the inputs from which the systematics weights are derived (for the BOSS CMASS sample these were stellar density and seeing, no such weights were applied for LOWZ \cite{Reid16}).  The measured correlation is about an order of magnitude lower than the cross-correlation signal between galaxy density and $\kappa$, so any error in the weights would have to be very significant to make a large impact on our results.

As a last test we cross-correlated a map of extinction \cite{SFD} against each of our galaxy overdensities in the NGC and the Planck $\kappa$ map.  Since the extinction map used to perform magnitude corrections is tracing both galactic and extragalactic structure \cite{Lenz:2017djx,Yahata:2006mf}, it is possible that incorrect extinction corrections may cause artificial correlation between galaxy overdensity and $\kappa$.  Assuming the projected galaxy over-density receives an additive contribution proportional to the extinction, we find that the galaxy autospectrum receives a bias due to extinction that should be well-below the percent level, in agreement with the finding of the regression analysis by the BOSS team that established no correlation of pixelized galaxy density with extinction value (and hence no need to include extinction as a contributor to the angular systematics weights \cite{Ross17,Reid16}), though those tests were done on the LOWZ and CMASS samples individually not the shuffled \textbf{z1} and \textbf{z3} samples. On the other hand, while our measurements are noisy, we find that the galaxy-$\kappa$ cross spectrum could be biased by up to a few percent from the extinction component in the observed galaxy field alone, even before taking into account the effect it has on the lensing estimator. If present, such a correction would constitute a significant fraction of our error budget, since our mock tests show that the \textbf{z3} bin alone should give us close to $5\%$ constraints on $\sigma_8$. However, we also find that our results are largely insensitive to changing the MV $\kappa$ map for the SZ-deprojected map, which should have larger contributions from galactic emission and CIB and thus suggests that any such bias is small. In any case, while such a bias would still be subdominant to our statistical uncertainty in this work, this test demonstrates that cross-correlation analyses can require more stringent foreground mitigation than each experiment individually.

\subsection{$\Lambda$CDM Constraints from Full Sample }

\begin{table}
    \centering
    \input{Tables/results}
    \caption{Cosmological constraints from fitting the full BOSS RSD+BAO data, with and without cross correlations with CMB lensing from Planck in the Northern Galactic Cap. The corresponding constraints from Planck are shown as comparison.
    }
\label{tab:lcdm_constraints}
\end{table}

\begin{figure}
    \centering
    \resizebox{\columnwidth}{!}{\includegraphics{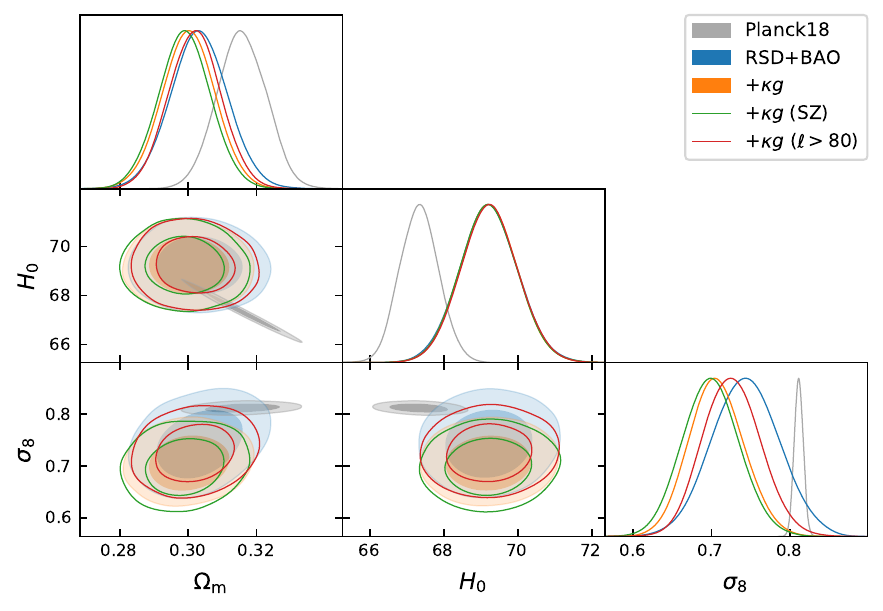}}
    \caption{The marginalized posteriors for the cosmological parameters from our analyses, compared to Planck (grey contours).  The blue shaded contours show constraints including only the RSD and BAO data, the other contours include the CMB lensing cross-correlation.  The orange contours include the full range of $C_\ell^{\kappa g}$, the green contours show the effect using the SZ-deprojected Planck lensing map and the red contours illustrate the effect of dropping the lowest $\ell$ point $C_\ell^{\kappa g}$.  }
    \label{fig:corner}
\end{figure}

\begin{figure}
    \centering
    \resizebox{\columnwidth}{!}{\includegraphics{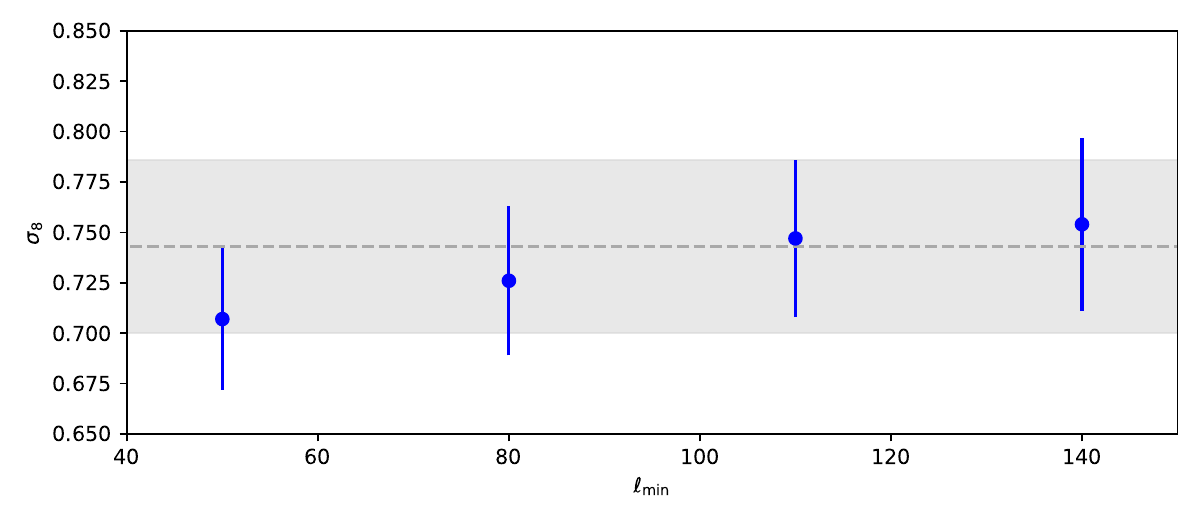}}
    \caption{Our marginalized constraints on $\sigma_8$ as a function of the minimum $\ell$ included in the lensing-galaxy cross-correlations, $\ell_{\rm min}$.  Note that we quote the minimum $\ell$, which starts at $\ell=50$, rather than the band center which would be larger by $\Delta\ell/2=15$.  The shaded grey band with dashed line shows the result from just the BAO+RSD data. }
    \label{fig:ellmin}
\end{figure}

Table~\ref{tab:lcdm_constraints} and Figure~\ref{fig:corner} shows the cosmological constraints obtained using the fiducial setup described in the previous sections, which includes RSD and BAO from the full BOSS sample, with and without additional information from cross-correlations with CMB lensing in the Northern galactic cap, as compared to constraints from Planck. The redshift-space only results are essentially identical to those recovered in ref.~\cite{Chen22} and while lower in amplitude are consistent with Planck constraints; we refer the reader to that work for further discussion of the information contained within RSD-only fits. As was seen in our mock analysis, adding in $C^{\kappa g}_\ell$ mainly serves to to tighten constraints on the amplitude $\sigma_8$, with slight improvement in the $\Omega_m$ constraint as well due to degeneracy breaking. Including lensing cross correlations in our analysis also decreases the mean $\sigma_8$ to $0.707 \pm 0.035$, in roughly $3\,\sigma$ tension with Planck and close to $1\,\sigma$ below the constraints from redshift-space alone. In terms of the $S_8 = \sigma_8 (\Omega_m/0.3)^{0.5}$ parameter best-probed by weak lensing, our analysis finds $S_8 = 0.707\pm 0.037$, compared to $S_8=0.747\pm 0.047$ from redshift-space data alone. Adding in lensing data leads to fractional improvements in the $S_8$ constraint greater than improvements in the $\sigma_8$ constraint due to degeneracy breaking; however, it is worth noting that even after including $C^{\kappa g}_\ell$ our $\sigma_8$ and $\Omega_m$ constraints remain slightly positively correlated due to the relative dominance of the redshift-space data. Future surveys where the lensing-galaxy cross correlation can be better measured should lead to further degeneracy breaking and further narrow constraints on the shape ($\Omega_m, H_0$) of the power spectrum by better measuring its amplitude ($\sigma_8$).

We can perform a few simple tests within the fiducial setup to ensure the robustness of our analysis and data. Our results are almost unchanged if we swap out the fiducial lensing map for the tSZ-deprojected one, also provided by the Planck collaboration: $\sigma_8$ changes to $0.699\pm 0.036$, a $0.2\,\sigma$ shift. This is a valuable cross check because the amount of cosmic infrared background and galactic foregrounds in the tSZ-deprojected maps is expected to be larger than in the (default) minimum variance map. We can also restrict our constraints to smaller scales, or larger $\ell$, since our systematics checks showed nontrivial large-scale angular systematics in the galaxy maps. As shown in the red contour in Figure~\ref{fig:corner}, dropping the lowest $\ell$ bin shifts the $\sigma_8$ constraint upwards to $0.726\pm 0.037$, representing a more than $0.5\,\sigma$ shift. Shifts of this magnitude are not unexpected when dropping data points, simply due to statistical fluctuations, though we note that the change in constraining power from dropping this one point is relatively meager. Figure~\ref{fig:ellmin} shows the effect of removing lensing data up to some scale $\ell_{\rm min}$. Removing the lowest $\ell$'s reduces the statistical tension with the redshift-space, with mean $\sigma_8$ steadily rising with $\ell_{\rm min}$. Much of this shift is because the most statistically constraining $\kappa g$ data are the low $\ell$ values (due to a combination of observational and theoretical errors), and thus the joint fit becomes increasingly dominated by the RSD+BAO, with increased error bars to match.  However a part of the shift to larger $\sigma_8$ is due to the $\kappa g$ pulling upwards.

Our analysis in this work is relatively constrained to work primarily at large angular scales. We have been very conservative in our scale cuts when modeling $C^{\kappa g}_\ell$, fitting to the same implied $k_{\rm max}$ as the redshift-space data which exhibit far more onerous nonlinearities due to small-scale velocities like fingers-of-god. More importantly, the Planck $\kappa$ map is signal dominated only at the lowest $\ell$ that we fit, significantly limiting our ability to better constrain the onset of nonlinearities in the data. Better data from current and planned CMB surveys will significantly expand the available information towards high $\ell$, allowing us to break bias degeneracies and check for systematics by comparing constraints from large and small scales. In order to maximally leverage this new information we will need to either validate PT models to beyond the conservative scales we use in this work or, more ambitiously, extend our modeling to smaller scales using simulations-based techniques. A particular class of these techniques, the so-called ``hybrid EFT'' (HEFT) approaches \cite{Modi20,anzu21,Zennaro:2021bwy,Hadzhiyska:2021xbv}, show particular promise because they share an identical set of clustering parameters with LPT, to which they reduce on large scales, while employing N-body dynamics to accurately predict clustering to the halo scale through a resummation scheme based on LPT. By extending perturbative bias modeling into the regime where dynamical nonlinearities are non-negligible, HEFT has the potential to break bias degeneracies and significantly tighten cosmological constraints from lensing-galaxy cross correlations, as we discuss in more detail in Appendix~\ref{app:highk}.

\subsection{Consistency Tests}
\label{ssec:consistency}

Our main result --- $\Lambda$CDM constraints from the combination of two- and three-dimensional data --- is not only in strong tension with Planck, but also in some tension with constraints from redshift-space data only. Our goal in this subsection is to investigate the source of this tension through considering subsamples of the BOSS data and by testing the consistency of amplitudes between RSD and lensing.

\begin{figure}
    \centering
    \resizebox{0.45\columnwidth}{!}{\includegraphics{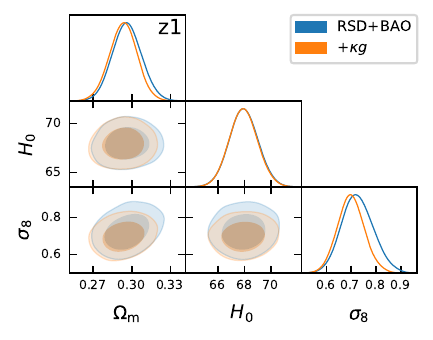}}
    \resizebox{0.45\columnwidth}{!}{\includegraphics{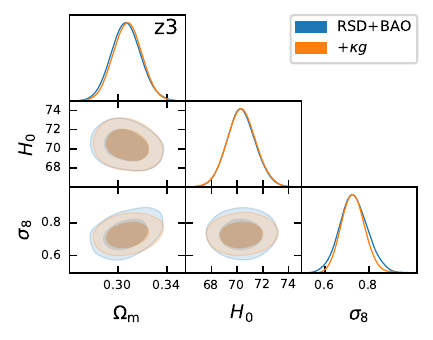}}
    \caption{The marginalized posteriors for the cosmological parameters from our analyses of the \textbf{z1} (left) and \textbf{z3} (right) samples.  The blue contours show the results using just the redshift-space data (i.e.\ BAO+RSD) while the orange contours include the galaxy-lensing cross-correlation. }
    \label{fig:corner_zs}
\end{figure}

Fig.~\ref{fig:corner_zs} shows the marginal posteriors for the cosmological parameters, with and without lensing, for the two BOSS redshift slices \textbf{z1} and \textbf{z3}. Within each subsample, the redshift-space data (including BAO) tightly constrain $\Omega_m$ and $h$ while the lensing data mainly sharpen constraints on $\sigma_8$. Comparing results with and without lensing, we see that $\sigma_8$ is more-or-less consistent with and without lensing in $\textbf{z3}$, but that the lensing data in $\textbf{z1}$ prefer lower $\sigma_8$ than RSD and BAO alone, producing visible shifts in both $\Omega_m$ and, more significantly, in $\sigma_8$. It is worth noting that redshift-space only constraints on $\sigma_8$ are highly consistent across redshift bins (see ref.~\cite{Chen22}) --- these results therefore suggest that the downward shift in $\sigma_8$ with the addition of lensing data are being driven chiefly by the $\textbf{z1}$ sample.

\begin{figure}
    \centering
    \resizebox{\columnwidth}{!}{\includegraphics{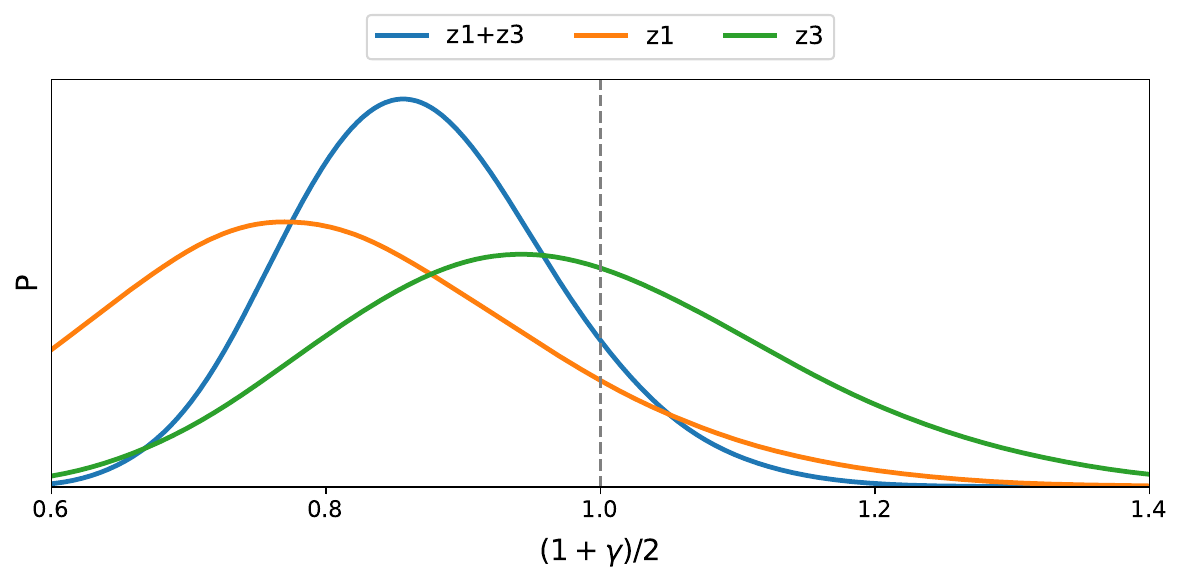}}
    \caption{Marginal posterior on the parameter, $c_\kappa=(1+\gamma)/2$, by which the galaxy-lensing cross-spectra are scaled in models with gravitational slip (\S\ref{sec:slip}).  The blue line shows the combined constraint from the high and low redshift samples while orange and green lines show \textbf{z1} and \textbf{z3}, respectively.  All of the constraints are consistent with the GR prediction of $\gamma=1$, though the lower redshift sample has lower $C_\ell^{\kappa g}$ than expected at modest significance.  We put a prior that $0.2<\gamma< 1.8$, so values of $c_\kappa$ below 0.6 are not allowed.  The \textbf{z1} sample hits this prior at the low end.  }
    \label{fig:gamma}
\end{figure}

As a further test, we can free the lensing amplitude $c_\kappa = (1+\gamma)/2$ from its prediction within general relativity and constrain it directly from the data. Measuring $c_\kappa$ acts both as a test of general relativity through measuring the gravitational slip $\gamma$ and as a consistency test between the redshift-space and lensing data. Figure \ref{fig:gamma} shows the marginal posterior on $c_\kappa$. The blue line shows the combined constraint from the high and low redshift samples while orange and green lines show \textbf{z1} and \textbf{z3}, respectively.
The combined-sample constraint is $\gamma=0.74^{+0.17}_{-0.21}$, with the \textbf{z1} sample giving $\gamma=0.66^{+0.17}_{-0.37}$ and \textbf{z3} giving $\gamma=0.94^{+0.28}_{-0.36}$. In line with our $\sigma_8$ results from the redshift subsamples, $c_\kappa$ constraints from \textbf{z3} are consistent with the prediction from general relativity, while those from \textbf{z1} show a mild preference for lower values, with a peak approximately $25\%$ below unity. This implies that the lensing-galaxy cross correlation in the latter sample is roughly $25\%$ lower than might be expected from the redshift-space data within $\Lambda$CDM, consistent with expectations based on Figs.~\ref{fig:corner_zs} and \ref{fig:ppd_lensing}. Nonetheless, our findings for both redshift slices and the combined sample are broadly consistent with the general-relativistic prediction of $\gamma= 1$, though they are again suggestive that lensing in the lower-redshift slice is driving our low $\sigma_8$ constraint. \edit{It is worth noting that, while it is possible to suppress the weak lensing amplitude relative to RSD via massive neutrinos, the mean suppression for our combined sample would translate (\S\ref{sec:slip}) to a neutrino mass fraction $f_\nu$ of roughly $30\%$, corresponding to $M_\nu \approx 4 $ eV, well above limits set by ground-based experiments.}

\subsection{Comparison to Previous Results: When they go low, we go...}
\label{ssec:lower}

\begin{figure}
    \centering
    \resizebox{\columnwidth}{!}{\includegraphics{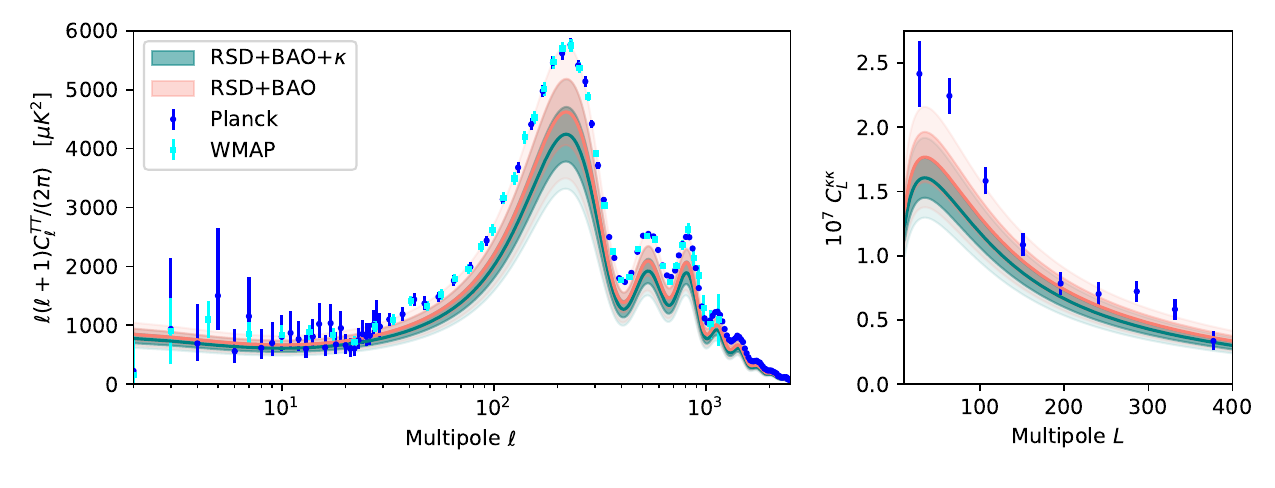}}
    \caption{The predictions for the $\Lambda$CDM model conditioned on our full data set (green; RSD+BAO+$\kappa$) or without the lensing (red; RSD+BAO) compared to the CMB angular power spectra measured by Planck \protect\cite{PlanckParams18} and WMAP \protect\cite{WMAP9} (left) or the convergence auto-spectrum ($C_L^{\kappa\kappa}$; right).  The shaded bands show the mean and $\pm 1$- and $\pm 2$-standard-deviation range for the model predictions while the points with errors show the best-estimate, foreground cleaned temperature or convergence power spectra from the CMB satellite missions.    }
    \label{fig:ppd_cmb}
\end{figure}

Our results add to the growing number of measurements at ``low $z$'' that have less clustering than inferred by Planck within the context of $\Lambda$CDM.  This is typically summarized in terms of $S_8=\sigma_8(\Omega_m/0.3)^{0.5}$.  In terms of this statistic we find $S_8=0.707\pm 0.037$ for the combined sample, lower than Planck's $S_8=0.832\pm 0.012$. To further illustrate the tension, Figure~\ref{fig:ppd_cmb} compares predictions for the CMB temperature and lensing anisotropies conditioned on our fiducial cosmological constraints and our redshift-space-only constraints compared to data from WMAP and Planck. Even when only redshift-space data are included the models with high likelihood underpredict both CMB statistics, and adding in cross correlations with lensing puts the best-fit models in strong tension with the CMB both by lowering the mean amplitude and tightening constraints.

We are not the first to study the combination of CMB lensing from Planck and galaxy clustering from BOSS. A number of authors have investigated cross correlations between the 2D (projected) galaxy clustering with lensing. Among the earliest was ref.~\cite{Pullen16}, who found within the best-fit Planck 2013 cosmology that the CMASS-lensing cross correlation amplitude was $0.754 \pm 0.097$ times the expected value. Ref.~\cite{Singh17} studied the galaxy-galaxy and galaxy-CMB lensing cross correlations using the BOSS LOWZ and CMASS samples assuming the Planck 2015 \cite{Planck2015} cosmology, finding correlation coefficients of $r_{cc} = 1.0 \pm 0.2$ and $0.78 \pm 0.13$, respectively, on scales with projected radii larger than $20 \Mpc$; in addition, cross-correlating with galaxy shears from the Sloan Digital Sky Survey they found that the amplitude of CMB lensing is reduced by a factor $A = 0.63 \pm 0.18$ below angular separations roughly corresponding to radial distances of $100 \Mpc$.

Varying cosmological parameters, ref.~\cite{Doux18} investigated the cross correlations of both BOSS galaxies and quasars, again finding that analyzing only the relatively low redshift galaxy-galaxy and galaxy-lensing cross correlations yields lower power spectrum amplitudes $\ln(10^{10} A_s)$, with a mean of roughly $2.9$, than when the CMB-lensing autospectra, which predominantly probe matter clustering at $z \gtrsim 2$, are included, in which case the derived amplitudes are consistent with Planck\footnote{We have inferred these numbers from the Figure 14 of ref.~\cite{Doux18} since no tables with constraints for each of these data combinations was provided.}. It should be noted that the low-redshift constraint includes the (relatively) higher redshift BOSS quasars, whose cross-correlation amplitude with CMB lensing more closely matches the Planck prediction than either LOWZ and CMASS; should the quasar data be dropped the galaxy-galaxy and galaxy-lensing data would presumably prefer even lower $\sigma_8$. Similarly, ref.~\cite{Singh20} analyzed the cross-correlation with LOWZ and CMASS and constrained the combination $\sigma_8^{0.8} \Omega_m^{0.6}$ to be $0.9 \pm 0.12$ times that predicted by Planck for both samples.  None of the BOSS and Planck $\kappa$ analyses above adopt the full set of bias and dynamical contributions to galaxy clustering required by fundamental symmetries as we do in this work and therefore do not exhaustively account for the possible contributions to clustering in the quasilinear regime --- they thus extract their amplitude information from a different set of scales, with greater theoretical uncertainty; however they are nonetheless suggestive (with relatively low significance) of a deficit in cross-clustering power between lensing and galaxy clustering at low redshifts when compared to Planck due to either unknown physics or systematics, as our more complete analysis finds at roughly $3\,\sigma$ significance.

Previous authors have also studied the combination of three-dimensional BOSS galaxy clustering and lensing analyzed in this work. These works have typically employed the so-called $E_G$ statistic, a test of general relativity proposed in ref.~\cite{Zhang07}. In that work, the linear theory of matter and galaxy clustering are combined with general-relativistic considerations to relate the ratio of galaxy-lensing cross correlations and redshift-space clustering anisotropies to fundamental quantities; schematically,
\begin{equation}
    \hat{E}_G \sim \frac{C^{\kappa g}_\ell}{P_2}, \quad \avg{\hat{E}_G} = \frac{(1+\gamma) \Omega_m}{2 f(z)} .
    \label{eqn:E_g}
\end{equation}
Assuming $\Omega_m$ is known, measuring this ratio in galaxy-lensing cross correlations constrains the gravitational slip, as we have done above. Previous works leveraging BOSS redshift-space clustering and CMB lensing arrived at mixed results; ref.~\cite{Pullen16} found lensing to be $2.6\,\sigma$ lower than that predicted for CMASS while ref.~\cite{Singh19} found both CMASS and LOWZ to be in excellent agreement with general relativity. It should be noted however that, unlike in our approach, using $E_G$ to constrain gravity requires working within linear theory with scale-independent bias --- the state-of-the-art in analyses such as ref.~\cite{Singh19}, who compare compute this ratio using a combination of BOSS galaxies, CMB lensing and cosmic shear surveys, account for the neglected gravitational nonlinearities by calibrating to simulations. In addition, the $E_G$ statistic is defined to be a single number computed by combining summary statistics of galaxy clustering and lensing evaluated at different scales and redshifts, reliant on the acceptability of the linear-theory prediction at a single redshift across these scales and redshifts in order to be compared to Equation~\ref{eqn:E_g}. By comparison, our approach is able to constrain the (scale-independent) gravitational slip leveraging both linear and quasilinear scales while simultaneously marginalizing over cosmological parameters directly, confirming the result of ref.~\cite{Pullen16} that the lensing-galaxy cross correlations measured from BOSS and Planck are lower than their observed redshift-space distortions imply, particularly for \textbf{z1}, though with only modest significance.

\begin{figure}
    \centering
    \resizebox{\columnwidth}{!}{\includegraphics{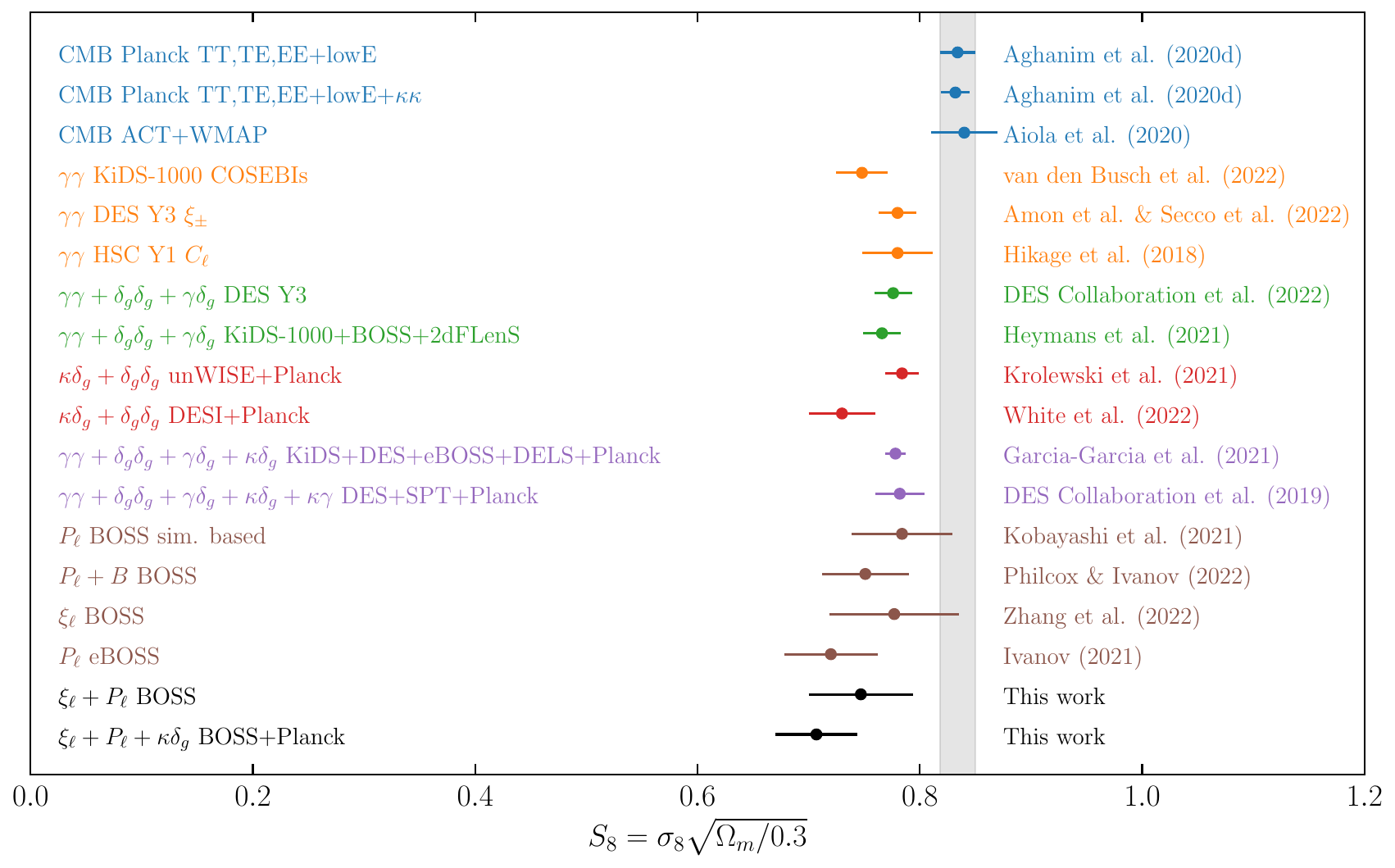}}
    \caption{A summary of recent $S_8$ constraints. The different colored points indicate different combinations of data that have been used in the constraints. In particular, we include constraints from the CMB (blue), cosmic shear (orange, $\gamma\gamma$), projected galaxy clustering and galaxy-galaxy lensing (green; $\delta_{g}\delta_{g}+\gamma\delta_{g}$), projected galaxy clustering and CMB lensing (red; $\delta_{g}\delta_{g}+\kappa\delta_{g}$), a combination of all of these (purple), redshift space clustering in various forms (brown), and the combination of data used in this work (black). As in Figure~\ref{fig:sigma8_sum}, we have limited ourselves to analyses using large scales. Despite the different models and statistics being used in these analyses, they all yield constraints below those from the CMB.}
    \label{fig:s8_sum}
\end{figure}

Beyond those combining BOSS and Planck there have been a wealth of recent results obtaining cosmological constraints from weak lensing and its cross correlation with galaxy clustering, most of which find $S_8$ to be lower than Planck but higher than that implied by our analysis, as shown in Figure~\ref{fig:s8_sum}. In the case of weak lensing only, the DES Y3 shear-only correlation function \cite{Amon21,Secco21} and harmonic space analyses \cite{Doux2022} find $S_8=0.772\pm0.017$ and $S_8=0.784\pm0.026$ respectively, $2\,\sigma$ lower than Planck but also in tension with our fiducial constraints at the $2\,\sigma$ level, though the tension is slightly reduced if we instead compare to the fiducial scale cut results instead of the $\Lambda$CDM optimized setup. A recent analysis of the KiDS-1000 data \cite{vandenBusch22} similarly found $S_8 = 0.748^{+0.021}_{-0.025}$, in slightly more tension with Planck but slightly closer to our result. Earlier analyses of cosmic shear in HSC \cite{Hikage:2018qbn} and CFHTLenS \cite{Heymans:2013fya} paint a similar picture. Adding in galaxy clustering from non-BOSS surveys, the DES Y3 ``$3 \times 2$'' analysis finds $S_8 = 0.776\pm 0.017$ \cite{DESY3} and, dropping the weak lensing autocorrelation,  a cross-correlation of unWISE-selected galaxies with Planck lensing \cite{Krolewski21} found $S_8 = 0.784\pm 0.015$, while using luminous red galaxies selected from DECALS ref.~\cite{White22} found $S_8=0.73\pm 0.03$. Our constraints are in modest ($\lesssim 2\,\sigma$) tension with most of these cross-correlations analyses except for this last work, for which $S_8$ is just shy of $1\,\sigma$ higher than our result. A combined analysis of cosmic shear, CMB lensing and galaxy clustering data, mostly sensitive to growth between $0.2 < z < 0.7$, by ref.~\cite{Garcia21} found $S_8 = 0.7781 \pm 0.0094$. The combination of these previous results (many of which probe similar redshift ranges to this work) could be an indication that the BOSS galaxy and Planck CMB $\kappa$ cross correlation measurement may be contaminated by some yet-unidentified foreground or systematic, since in the absence of such an effect we would be probing similar epochs of structure formation, though more concrete conclusions regarding these tensions will likely have to wait for upcoming CMB lensing measurements from e.g.\ ACT, whose instrument noise on the scales we study will be significantly reduced.

\begin{figure}
    \centering
    \resizebox{\columnwidth}{!}{\includegraphics{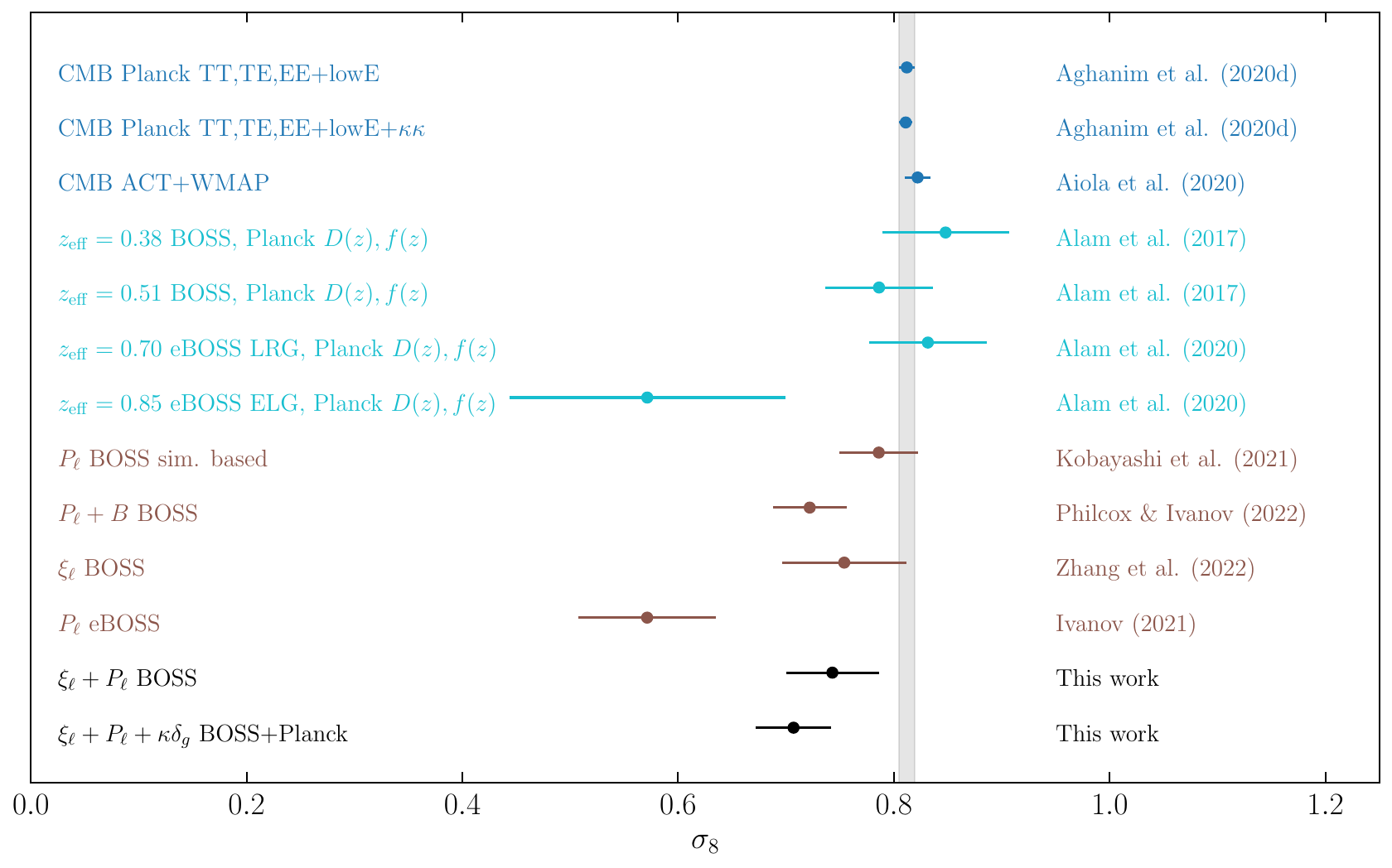}}
    \caption{A summary of $\sigma_8$ constraints from recent CMB measurements (blue) compared to those made using template based (cyan) and full-shape (brown) fits to anisotropic redshift space correlation functions ($\xi_{\ell}$), power spectra ($P_{\ell}$), and bispectra ($B$), as well as including $C^{\kappa g}$ in this work (black). For the purposes of this figure, we have limited ourselves to analyses that focus on large scales, although we provide a more complete overview in Section~\ref{ssec:lower}. For the template based fits, we quote ``consensus'' constraints, which are weighted averages of multiple analyses. These template based fits constrain $f\sigma_8(z_{\rm eff})$ directly, so we have assumed the best fit cosmology from \cite{PlanckParams18} to compute $D(z_{\rm eff})$ and $f(z_{\rm eff})$ in order to convert to $\sigma_8(z=0)$. The template based fits are largely more consistent with the CMB constraints, other than the eBOSS ELG point. The full-shape analyses yield lower $\sigma_8$ values than the template based fits and the CMB, and are relatively consistent despite using significantly different models and statistics. The inclusion of the CMB lensing data in our analysis tightens our $\sigma_8$ constraint by partially breaking the degeneracy between $\Omega_{\rm m}$ and $\sigma_8$, but also drives it significantly lower than our RSD-only fits.}
    \label{fig:sigma8_sum}
\end{figure}

The fact that $C^{\kappa g}_\ell$ has an amplitude close to $20\%$ lower than implied by redshift-space clustering hints that there may be an unknown systematic leading to internal inconsistency within the data. The latter measurement is by now under excellent theoretical control and, in addition to our results, recent analyses of BOSS by refs \cite{Zhang21b, Philcox22} using the redshift-space galaxy 2-point function in configuration and Fourier space give $\sigma_8 = 0.766\pm 0.055$ and $0.737^{+0.040}_{-0.044}$\footnote{We have adopted their constraints using the public power spectra modeled with fixed $n_s$, for better comparison with our analysis.} both in excellent agreement with our results, as shown in Figure~\ref{fig:sigma8_sum}. These recent analyses employ improved models of galaxy clustering compared to the earlier (official) results of the BOSS collaboration, marginalize over cosmological parameters like $\Omega_m$ and $H_0$ beyond the growth rate $f(z)$, and also correct for errors in the window-function normalization. Together, this new generation of BOSS constraints confirms that there is a deficit of power in $C^{\kappa g}$ compared to that inferred from the velocity-induced anisotropy in galaxy clustering\footnote{The $\sigma_8$ constraints from eBOSS ELG's in refs.~\cite{Ivanov:2021zmi,Alam20} are notably lower than the others shown in Figure~\ref{fig:sigma8_sum}, but the strong tension with other measurements at similar redshifts, including eBOSS LRG's, suggest that this may be due to systematics (e.g.~the large redshift range fit).}, though at lower significance given that redshift-space constraints on $\sigma_8$ are considerably weaker due to bias degeneracies. Unlike in the case of weak lensing, however, redshift-space analyses like the above are able to independently constrain parameters like $\Omega_m$ and $\sigma_8$ instead of highly degenerate combinations like $S_8$. It is worth noting that analyses of BOSS galaxy clustering using N-body based emulators \cite{Kobayashi21,Zhai2022,Yuan2022}, or similar simulation based techniques \cite{Lange2022}, also return constraints very close to our redshift-space result, with smaller error bars, though we caution that these constraints rely on far more restrictive assumptions about the small-scale behavior of galaxy clustering and thus have a larger systematic error. 

A more theoretically robust alternative for improving cosmological constraints from galaxy clustering is to also perturbatively model higher n-point functions; when the bispectrum is taken into account, ref.~\cite{Philcox22} find that their $\sigma_8$ constraint tightens to $\sigma_8 = 0.722^{+0.032}_{-0.036}$, a similar gain in constraining power to the addition of lensing information seen in this work. The bispectrum in principle breaks the $f\sigma_8$ degeneracy in galaxy clustering and can provide $\sigma_8$ information beyond that in the velocities; curiously, this nonlinear information also prefers (slightly) lower $\sigma_8$ than the linear RSD alone. \edit{In discussing ref.~\cite{Philcox22} here and above we have used their results with the spectral index $n_s$ fixed to better match the analysis setup employed in this work---freeing $n_s$ in our analysis yields verys similar redshift-space only constraints to that work, while adding in lensing data lowers $\sigma_8$ by about $1\sigma$ as in the fixed $n_s$ case, as we show in Appendix~\ref{app:ns}.}  Higher-order statistics and cross correlations with nonlinear matter through lensing yield competitively tight constraints on cosmological parameters, and will provide complimentary clustering information in upcoming surveys useful both as internal consistency checks and probes of new physics beyond the standard, linear redshift-space distortions traditionally probed by spectroscopic surveys.

\section{Conclusions}
\label{sec:conclusions}

The two and three dimensional clustering of galaxies measured by spectroscopic surveys offer complementary cosmological information: the latter encodes the shape of the primordial power spectrum, distance information through baryon acoustic oscillations, and cosmic velocities through redshift-space distortions, while the former, when in combination with probes of weak lensing like the CMB, probes the amplitude of matter fluctuations through their induced Weyl potential. In this paper we lay out a formalism to jointly analyze these two distinct probes in the language of effective perturbation theories, presenting a proof-of-principle analysis using Lagrangian perturbation theory to model publicly available data from galaxies in the BOSS survey \cite{Dawson13,BOSS_DR12} and CMB lensing data from the Planck satellite \cite{PlanckLens18}.  To our knowledge this is the first such joint analysis to use a consistent theoretical model valid into the quasilinear regime taken all the way to the data (2-point functions), rather than utilizing linear theory and compressed statistics derived from it. This is significant because perturbation theory allows for rigorous and systematic modeling of structure formation on large scales scales with minimal theoretical assumptions and will be invaluable to distinguish true cosmological signals from either theory or data systematics for current and upcoming surveys.

A particular goal of this work has been to set up this analysis in a theoretically well-motivated way (\S\ref{sec:theory}). To this end we have, for example, been careful in our perturbative treatment of neutrinos, which affect galaxy-galaxy and galaxy-matter spectra in meaningfully different ways, and we introduced redshift-dependent weights to the galaxy-lensing cross-correlations measurements to ensure they probe clustering at the same effective redshift as the three-dimensional power spectrum. As a test of our formalism, we validate our various theoretical choices and approximations using lightcone mocks of BOSS galaxies (\S\ref{sec:mocks}) based on the Buzzard simulations \cite{DeRose2019}, showing that our model is able to recover the ``truth'' to within the statistical scatter expected from the volume of these simulations (\S\ref{ssec:mock_results}).

The data consist of 1,198,006 galaxies covering 25\% of the sky (10,252 sq.deg.) \cite{Reid16}, and the Planck lensing map covering approximately $60\%$ of the sky, though for cross-correlation with the Planck lensing maps we utilize only the 7,143 sq.deg.\ in the NGC. The Planck lensing map is signal dominated near $\ell\approx 40$ \cite{PlanckLens18}. We use the low- (\textbf{z1}; $0.2<z<0.5$) and high-redshift (\textbf{z3}; $0.5<z<0.75$) samples based on spectroscopic redshifts as defined by the BOSS collaboration \cite{Alam17}. As also discussed in ref.~\cite{Chen22}, while making new galaxy samples and measurements more tailored to our analysis is in principle possible, doing this work --- including re-making enough mock measurements to estimate the covariance matrix --- would require resources beyond the scope of this project. We therefore leave data-side optimization of this analysis to future work.

The main results of our analysis, constraints on $\Omega_m$, $H_0$ and $\sigma_8$ based on the combination of BOSS galaxy clustering and Planck CMB lensing, are described in \S\ref{sec:results}. We perform systematics tests of the galaxy and lensing maps in \S\ref{sec:systematics}, finding that the systematics weights for the BOSS galaxies would have to have left significant traces of the mitigated systematics in the maps to have even few-percent effects on the cross-correlation amplitude, $C^{\kappa g}_\ell$. Cross correlating the galaxy and lensing maps with maps of extinction, an effect not included in the systematics weights for BOSS galaxies due to its relatively small effect, indicates that extinction errors also have a small impact at the at-most few-percent level in cross correlations. We also cross-correlate the non-overlapping low and high redshift (\textbf{z1}, \textbf{z3}) samples, finding spurious large scale correlations in the lowest $\ell$ bins and in the SGC --- out of an abundance of caution we therefore drop these data points from our main analysis.

Our main result, using the full three-dimensional galaxy clustering data from BOSS and CMB lensing in the NGC, is summarized in Table~\ref{tab:lcdm_constraints} and Figure~\ref{fig:corner}. While the three-dimensional clustering data including power spectra and reconstructed correlation functions strongly constrain $\Omega_m$, $H_0$ through the shape of the linear power spectrum, including lensing information through $C^{\kappa g}_\ell$  sharpens the amplitude ($\sigma_8$) constraint by roughly $20\%$ and, since lensing probes this amplitude multiplied by the matter density, also somewhat sharpens the constraint on $\Omega_m$. Adding the lensing data, which are substantially lower on large scales than the redshift-space data might predict (Fig.~\ref{fig:ppd_lensing}), has the effect of lowering both, though $\Omega_m$ decreases by less than half a sigma and our model still predicts an acoustic scale ($\sim \Omega_m h^3$) highly consistent with the narrow range allowed for by Planck. On the other hand, including lensing we constrain $\sigma_8 = 0.707 \pm 0.035$, in roughly $3\,\sigma$ tension with Planck constraints, and an implied lensing amplitude, $S_8$, roughly $2\,\sigma$ lower than cosmic shear analyses, though in good agreement with another effective-theory based analysis of BOSS galaxy clustering including the bispectrum. 

Looking at subsamples of our data separately we find that the drop in $\sigma_8$ is driven primarily by the low redshift sample \textbf{z1} (\S\ref{ssec:consistency}). By freeing the gravitational slip $\gamma$, we find for that sample that the implied ratio of the Weyl to Newtonian potentials $c_\kappa = (1+\gamma)/2$ is roughly $20\%$ lower than predicted by general relativity, but at less than $2\,\sigma$ significance (Fig~\ref{fig:gamma}), and indeed we do not detect any deviation from unity for this ratio at more than $2\,\sigma$ significance in either the redshift slices separately or in combination. It is worth noting that our ability to leverage the relative amplitudes of galaxy clustering and galaxy-lensing cross correlations has implications beyond gravitational slip. For example, massive neutrinos will tend to suppress the latter relative to the former, though at a level far below the current level of constraints. Conversely, much recent attention has been paid to whether selection-induced anisotropic bias can be a significant contaminant of the RSD signal \cite{Hirata09,Obuljen20}. Such an effect would add a term $b_{zz} s_{zz}$, where $s_{zz}$ is the component of the shear tensor along the line-of-sight, to the bias expansion (at leading order) such that the linear galaxy overdensity in redshift-space becomes 
\begin{equation}
  \delta_{g,s}(\bk,z) = \left[ 1 + b_1 - \frac{b_{zz}}{3} + \left(f(z) + b_{zz}\right)\mu^2 \right]\ \delta_m(\bk) + \cdots ,
\end{equation}
leading to an exact degeneracy between $b_{zz}$ and the amplitude of the RSD anisotropy, and to biases in values of $\sigma_8$ inferred from redshift surveys. However, this degeneracy can be broken by the inclusion of lensing cross correlations, which measure $\sigma_8$ through the $\mu\approx 0$ component; together with spectroscopic clustering measurements, which also determine $\Omega_m$ and thus $f(z)$, this combination allows for a clean measurement of $b_{zz}$ and $\sigma_8$. Indeed, since Figure~\ref{fig:gamma} implies that lensing and redshift-space clustering amplitudes are roughly in agreement, $b_{zz}$ at least cannot be of order unity for either redshift slice. Future surveys will significantly improve our ability to exploit this synergy between lensing and RSD.

While our results are sufficiently constraining to considerably sharpen the tension in $\sigma_8$ between the CMB and LSS in $\Lambda$CDM, we still remain limited by the data that we use and by our modeling. Luckily, we anticipate rapid progress in both directions in the very near future. The galaxy maps are already sample variance dominated on the large scales from which we derive most of our cosmological information, so the next major improvement in errors at intermediate $\ell$ will come from CMB lensing maps with lower noise than Planck. Redshift-space clustering measured from DESI will also dramatically improve over those we used here.  Using maps optimized for cross-correlations, with careful attention to foreground cleaning or hardening, derived from more sensitive and higher angular resolution ground-based experiments will dramatically lower the uncertainties of $C_\ell^{\kappa g}$. \edit{These lower-noise measurements should allow us to better distinguish between the shapes of various nonlinear contributions to $C^{\kappa g}_\ell$ even on the scales we have analyzed in this work. More ambitiously, as discussed in Appendix~\ref{app:highk}, recent work extending the LPT modeling in real space to more nonlinear scales using hybrid N-body models \cite{Modi20} can allow us to self-consistently double the $\ell$ reach of our formalism by switching the perturbative calculations of $C^{\kappa g}_\ell$ for an emulator (e.g. \cite{anzu21}).} Combined with improved modeling \edit{future experiments} will improve the constraints on the power spectrum amplitude, $\sigma_8$\edit{, and allow us to check the consistency between constraints derived from large and small scales, even from within the same and related theoretical models}. The well-motivated and tested theoretical framework outlined herein should be ideal for such future work.

In terms of improvements in the input maps, we note that reducing systematics at low $\ell$ is particularly important for improving constraints.  Scale-dependent bias and astrophysical effects become increasingly important at small scales (larger $k$) and having sufficiently constraining data over a range of scales is crucial for constraining departures from linearity and breaking degeneracies between bias and effective-theory parameters. While upcoming CMB experiments will straightforwardly reduce the noise in CMB lensing measurements on quasilinear scales (intermediate $\ell$), there is also much to be gained by ensuring that both the galaxy and CMB lensing data are uncontaminated on large scales.

\acknowledgments
We thank Simone Ferraro, Anton Baleato Linzancos and Noah Sailer for helpful conversations about CMB lensing and foregrounds, and suggestions for limiting systematic errors. We similarly thank Noah Weaverdyck for useful discussions of foregrounds in galaxy surveys, and Ashley Ross for enlightening discussions of potential systematics in the BOSS catalogs. We thank Zvonimir Vlah for continued discussions on perturbation theory.
M.W.\ thanks Uros Seljak for numerous conversations on cosmological modeling and inference and for comments on an early draft.
S.C.~is supported by the DOE.
M.W.~is supported by the DOE and the NSF.
J.D.~is supported by the Lawrence Berkeley National Laboratory Chamberlain Fellowship.
N.K.~is supported by the Gerald J.~Lieberman Fellowship. 
We acknowledge the use of \texttt{NaMaster} \cite{Alonso18}, \texttt{Cobaya} \cite{CobayaSoftware, Torrado21}, \texttt{GetDist} \cite{Lewis19}, \texttt{CAMB} \cite{Lewis00} and \texttt{velocileptors} \cite{Chen20} and thank their authors for making these products public.
This research has made use of NASA's Astrophysics Data System and the arXiv preprint server.
This research is supported by the Director, Office of Science, Office of High Energy Physics of the U.S.\  Department of Energy under Contract No.\ DE-AC02-05CH11231, and by the National Energy Research Scientific Computing Center, a DOE Office of Science User Facility under the same contract.

\appendix

\section{Neutrinos}
\label{app:pt_nus}

\begin{figure}
    \centering
    \includegraphics[width=0.8\textwidth]{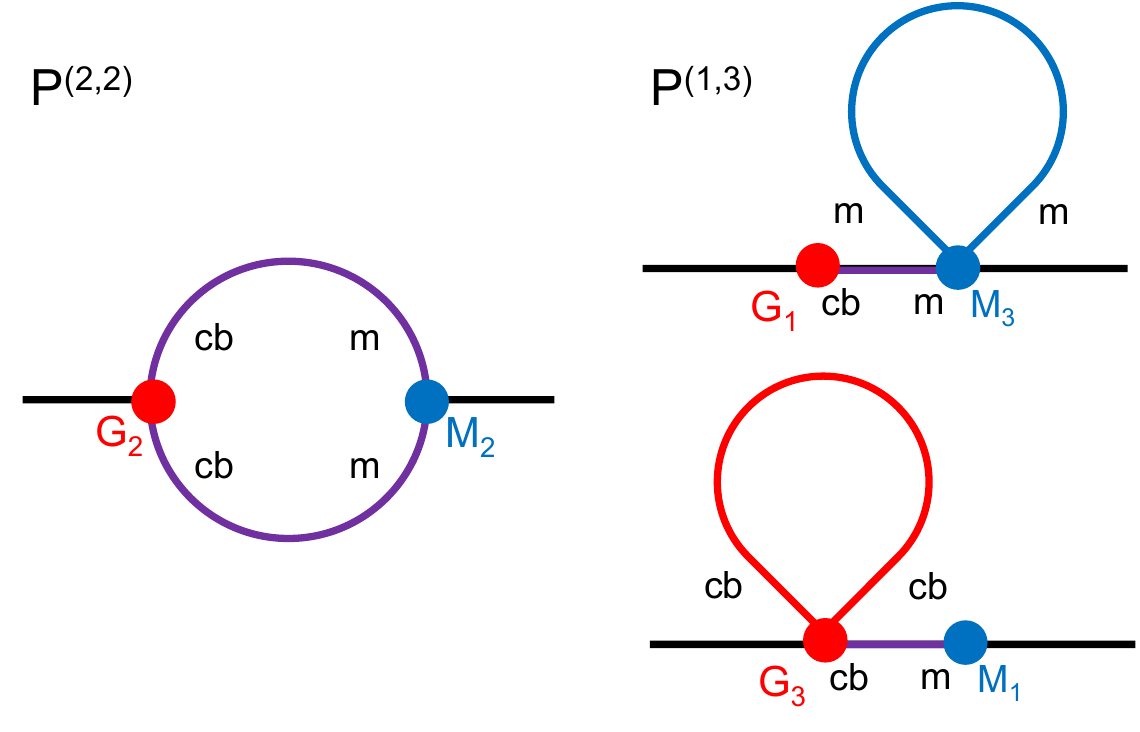}
    \caption{Diagrams of contributions to the 1-loop matter-galaxy power spectrum. Red vertices ($G_n$) and lines (linear spectra) correspond to galaxies and the ``cb'' component while blue ones ($M_n$) correspond to matter including neutrinos. Purple lines indicate the cb-m cross spectrum.}
    \label{fig:xcorr_diagrams}
\end{figure}

Throughout this paper we have worked within the approximation that galaxy clustering within massive neutrino cosmologies traces the dark matter-baryon component and that its autocorrelation can be modeled by computing perturbation-theory predictions using $P_{\rm cb,lin}$ as the input linear power spectrum. On the other hand, within the same approximation the nonlinear galaxy-matter cross correlation has to be computed via a mixture of loop integrals involving $P_{\rm cb,lin}, P_{\rm m,lin}$ and $P_{\rm cb, m,lin}$, which are shown in Figure~\ref{fig:xcorr_diagrams} as red, blue and purple lines, respectively. Roughly speaking, this is because contractions involving only galaxy (``cb'') or matter (``m'') vertices involve autospectra while contractions across vertices produce cross spectra.

While the computation of these mixed diagrams is in principle straightforward, in this paper we would like to make use of existing perturbation codes like \texttt{velocileptors} and therefore make use of the additional approximation that cross spectra can be computed using only $P_{\rm cb, m,lin}$ as the input power spectrum. As discussed in \S\ref{ssec:lin_cosmo}, this properly captures the shape of the transition from clustered to unclustered neutrinos at the free streaming scale. However, at the 1-loop level this prescription is not strictly correct---as shown in Figure~\ref{fig:xcorr_diagrams}, the $(2,2)$ contributions (left) depend on the cross $P_{\rm cb, m,lin}$ and will be correctly captured but the $(1,3)$ contributions involve cb-m autospectra, which differ from calculations using the cross spectrum by order $\mathcal{O}(f_\nu \PL)$.

Fortunately, the differences due to the above mistake will tend to be small for two reasons: the neutrino mass fraction $f_\nu$ is small and, since they only appear at the 1-loop level, they will be further suppressed relative to leading contributions, such that the total error will be of order $\mathcal{O}(f_\nu \PL^2).$ For example, the error incurred by evaluating the bottom-right diagram in Figure~\ref{fig:xcorr_diagrams} will be
\begin{equation*}
    3 M_1(\bk) P^{\rm lin}_{\rm cb,m}(\bk) \int_{\bp} G_3(\bk,\bp,-\bp) \big( P^{\rm lin}_{\rm cb,m}(\bp) - P^{\rm lin}_{\rm cb}(\bp) \big) \sim f_\nu P^{\rm lin}_{\rm cb,m}(\bk) \int_{|\bp|>k_{fs}} G_3(\bk,\bp,-\bp) P^{\rm lin}_{\rm cb}(\bp)
\end{equation*}
where we have used that, at wavenumbers larger than the free streaming scale $k_{fs}$, $P_{cb,m} \approx (1-f_\nu) P_{cb}$ and we have defined the galaxy third-order kernel $G_3$ and $\int_\bp = \int \frac{d^3\bp}{(2\pi)^3}$. A particular concern might be that the above mistake would lead to a contribution degenerate with the linear bias $\Delta b\ P^{\rm lin}_{cb,m}$, where $\Delta b$ would be sourced by dark matter density fluctuations above the $k_{fs}$, leading to inconsistencies in the bias definitions used in our real and redshift-space analyses; however, such a contribution due to short-wavelength (UV) modes can be prevented by adopting normal-ordered bias operators \cite{McDRoy09,Des18}, sometimes also referred to as renormalized operators, as is done in \texttt{velocileptors} \cite{Chen20,Chen21}. Beyond this the next leading correction will be of the form $f_\nu k^2 \Sigma^2 P^{\rm lin}_{cb}$, where $\Sigma^2$ is the variance of linear ($cb$) displacements from small-wavelength modes\footnote{We might also expect the cross-spectrum BAO damping paramater $\Sigma^2_{\rm BAO}$ to be modified by order $f_\nu$, but this will not concern us since the projected $C^{\kappa g}_\ell$ are insensitive to BAO.}. Since such a correction has to be subleading even without the factor of $f_\nu$ in order for perturbation theory to be valid, it will be highly suppressed and negligble for most purposes.

\section{Redshift-Dependent Galaxy Selection Effects and Cuts}
\label{app:wsys_z}

One possibility for the non-zero cross-correlation we observe between the \textbf{z1} and \textbf{z3} slices, whereas the similar correlation between the LOWZ and CMASS samples is smaller \cite{Doux18}, is that the systematics weights have a redshift dependence, either implicitly or induced by a redshift-dependence in some other property (size, luminosity, color, etc.).  Since such a dependence has been neglected in deriving and applying the weights this would imply that deriving the weights on one set of samples but then analyzing a shuffled set leads to correlations.

The BOSS collaboration derived the weights by which they correct for observational systematics by removing linear trends in the pixelized, projected galaxy number density vs.\ systematic template \cite{Reid16}.  He we provide a simple toy model for how redshift-dependence in the impact of systematics could lead to a bias in such a procedure.
Let the ``true'' redshift distribution of a galaxy sample be $dN/dz$.  Given a multiplicative selection bias, say due to some galactic foreground, the observed distribution is
\begin{equation}
    \frac{dN_{\rm obs}(\theta)}{dz} = \left[1 + T(\theta,z) \right]\ \frac{dN}{dz}
\end{equation}
where the angular dependence is entirely due to the foreground contamination. To correct for this we define a systematic weight, $w(\theta)$, whose $\theta$-dependence arises from the systematics template, such that the distribution of projected densities is uniform when plotted against the value of the systematic, i.e.
\begin{align}
    w^{-1}(\theta) &= \frac{1}{\bar{N}} \int dz\ \frac{dN}{dz}\  \big(1 + T(\theta,z) \big) \nonumber \\
    &\approx 1 + T(\theta,z_0) + \frac{1}{\bar{N}} \Big( \int dz\ \frac{dN}{dz}\   (z - z_0)\ \Big)\ T'(\theta,z_0)
\end{align}
where $\bar{N}$ is the average projected number density of the sample \textit{sans} systematics and we have Taylor-expanded the redshift dependence to first order. The final term vanishes when $z_0=z_{\rm mean}$. However, if the sample is now split up into further redshift bins $z \in (z_a,z_b)$ this will lead to an angular correction proportional to the offset in the mean redshift of the new sample compared to the old one:
\begin{align}
    N_{ab}(\theta) &= w^{-1}(\theta) \int_{z_a}^{z_b} dz\ \frac{dN}{dz}\  \big(1 + T(\theta,z) \big) \nonumber \\
    &= \frac{ \bar{N}_{ab} (1 + T_{0} + T_{0}^\prime\, \Delta z_{ab})  }{1 + T_{0}} \nonumber \\
    &\approx \bar{N}_{ab} ( 1 + T_{0}'(\theta) \Delta z_{ab})
    \quad\mathrm{with}\quad
    \Delta z_{ab} \equiv \frac{1}{\bar{N}_{ab}}\int_{z_a}^{z_b} dz\ \frac{dN}{dz}\   (z - z_{\rm mean}).
\end{align}

\section{Prospects for degeneracy breaking by pushing to smaller scales}
\label{app:highk}

\begin{figure}
    \centering
    \includegraphics[width=\textwidth]{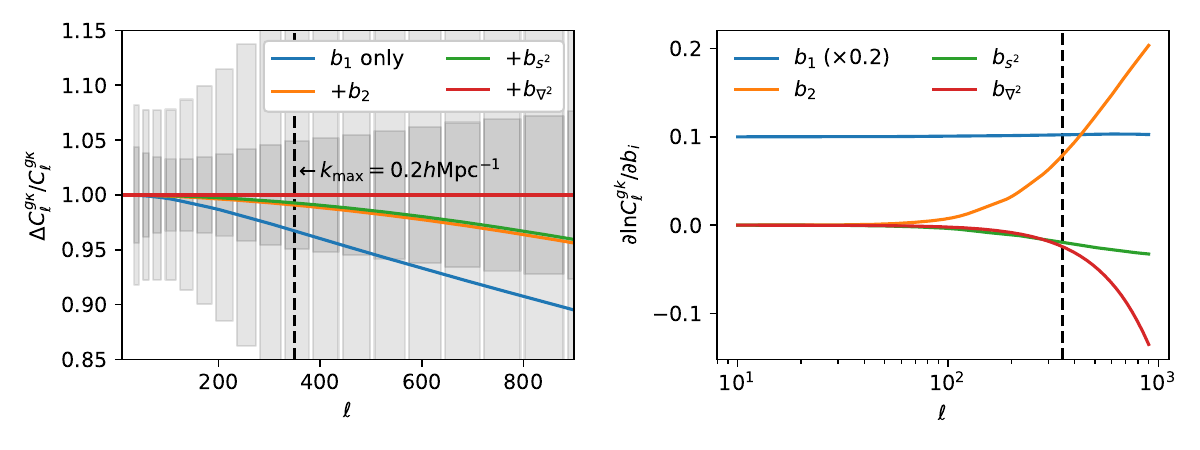}
    \caption{\emph{Left}: The impact of including higher order Lagrangian bias in the observable signal of $C_\ell^{g\kappa}$, compared to the uncertainty expected for a $z_{\rm eff} \approx 0.59$ DESI-like number density and lensing noise corresponding to the Planck and Simons Observatory (SO) surveys. \emph{Right:} The logarithmic derivatives of the $C_\ell^{g \kappa}$ power spectrum with respect to bias parameters as a function of scale. The vertical dashed line indicates the highest $\ell$ used in this analysis, which as discussed in \S~\ref{ssec:priors} corresponds to scales of $k \approx 0.2 h{\rm Mpc}^{-1}$. The highest $\ell$ shown corresponds to $k \approx 0.6 h{\rm Mpc}^{-1}$ for this sample, the smallest scales where we expect second-order hybrid LPT approaches to recover unbiased results. }
    \label{fig:highk_lever}
\end{figure}

A limitation of the techniques used in this paper arises from the fact that contributions from higher order biases (as well as shot-noise in the case of the auto-spectrum) are nearly degenerate on the quasi-linear scales we have probed in this work.  In Fig.~\ref{fig:ppd_lensing}, we can see the uncertainties associated with poorly constrained counterterms reduce the constraining power of scales which are well within the perturbative regime.  While the combination of probes we use depends on these parameters in different ways, these degeneracies are fundamentally present in the 3D spectra, $P_{gg}$ and $P_{gm}$, at these scales.

In Fig.~\ref{fig:highk_lever} we show the fractional change in including higher order bias operators in the model for $C_\ell^{g\kappa}$ for a DESI-like sample of galaxies cross-correlated with Planck and SO CMB lensing. The bias parameters used are derived from the field-level inference of \cite{Kokron21} for a DESI-like HOD. The binning adopted uses bins $\Delta \ell \approx 3\sqrt{\ell}$, and each error bar should be thought of as an independent data point. At scales $\ell > 350$, relative to the SO uncertainties, the impact of including each additional operator is significant. The sole exception is the tidal bias $b_{s^2}$, which is known to impact the $P_{gm}$ spectrum weakly. In the galaxy--$\kappa$ cross-spectrum we can also observe that for $\ell$s that probe $k<0.2 h{\rm Mpc}^{-1}$, many of the contributions are approximately degenerate, as shown in the right-hand panel of Fig.~\ref{fig:highk_lever}. However, at smaller scales the responses of observables like $C_\ell^{g\kappa}$ to changes in bias parameters become distinct, implying that accessing these smaller scales can help lift these degeneracies.

Despite the challenges of pushing to smaller scales in redshift space, the prospect of extending galaxy-$\kappa$ analyses to higher $\ell$ is tantalizing. This observable is largely insensitive to the imapact of RSDs and can be readily modelled by hybrid LPT techniques \cite{Modi20,anzu21,Zennaro:2021bwy,Hadzhiyska:2021xbv} that combine the same Lagrangian bias expansion (Eqn.~\ref{eqn:lagbias}) with N-body dynamics to give an accurate model of $P_{gm}$ up to $k_{\rm max} \simeq 0.6 \, h\, {\rm Mpc}^{-1}$, corresponding to $\ell_{\rm max} \approx 900$.  Extending our analysis to such $\ell_{\rm max}$, once the CMB lensing noise is sub-dominant at these scales, will allow for a significantly longer lever arm that will simultaneously break degeneracies between bias parameters, include more independent modes in the analysis, and reduce the impact of potentially systematics-dominated lower-$\ell$ modes. 

\section{Fits with Free Spectral Index \boldmath\texorpdfstring{$n_s$}{ns}}
\label{app:ns}

\begin{figure}
    \centering
    \includegraphics[width=0.8\textwidth]{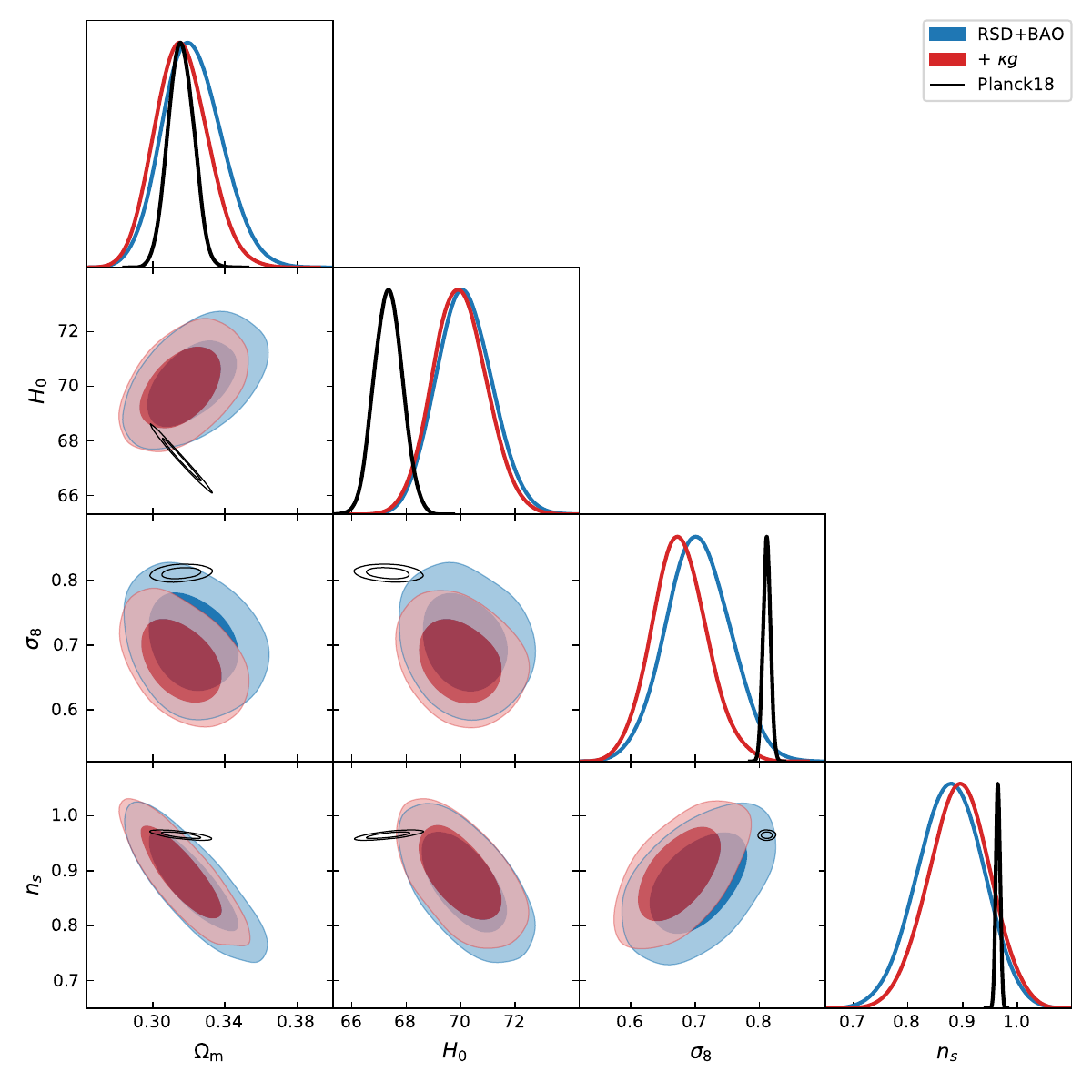}
    \caption{Marginalized cosmological posteriors as in Figure~\ref{fig:corner} for the case where $n_s$ is varied. Blue and red contours show results without and with lensing data. The corresponding posteriors from Planck are shown in black for comparison.}
    \label{fig:results_ns}
\end{figure}

As described in the body of the paper, our main analyses in this work have been performed by fixing the spectral index to the best-fit value for Planck \cite{PlanckLegacy18}. We have made this choice because the large-scale structure data we consider in this work are unable to robustly constrain $n_s$ and leaving it free results in the exploration of parts of the $\Lambda$CDM parameter space strongly ruled out by Planck. However, for completeness, and to more fully compare with other results in the literature, in this Appendix we consider the effect of extending our fits to include $n_s$ with an uninformative uniform prior $\mathcal{U}(0.5, 1.5)$.

The main results of this extended analysis are shown in Table~\ref{tab:results_ns} and Figure~\ref{fig:results_ns}. When only three-dimensional redshift-space galaxy data, including BAO, are considered, this analysis yields a mild ($\sim 1.5\sigma$) preference for values of $n_s$ below the Planck value, leading to noticeable shifts and widening in the $\Omega_m, H_0, \sigma_8$ posteriors --- notably, the mean $\sigma_8$ shifts further downward relative to the CMB --- though the redshift-space only posteriors are still in decent agreement with Planck, as the shifts in these parameters mostly lie along poorly constrained degeneracy directions\footnote{Indeed, we note that both the best-fit elements in the chain and found by {\sc Cobaya}'s in-built optimization routine have $n_s$ and $\sigma_8$ roughly $1\sigma$ above the mean and in agreement with the posteriors with $n_s$ fixed, suggesting a potential volume effect.} and the $2\sigma$ regions of BOSS and Planck overlap in all cases. These shifts in the RSD constraints are in excellent agreement with ref.~\cite{Philcox22}. This one-parameter extension does not alleviate the low $\sigma_8$ tension in the main analysis; and as in the main analysis, adding in cross correlations with CMB lensing leads to a further drop in $\sigma_8$ of roughly the same fractional size as that in our fiducial analysis ($\approx 5\%$) while the other $\Lambda$CDM parameters remain largely unchanged.

\begin{table}
    \centering
    \input{Tables/results_ns}
    \caption{Cosmological constraints from BOSS with and without CMB lensing when $n_s$ is varied.}
\label{tab:results_ns}
\end{table}

\bibliographystyle{JHEP}
\bibliography{main}
\end{document}

%% file: Tables/cosmo_priors.tex
\begin{tabular}{ || c || c | }
    \hline
    Parameter & Prior \\
    \hline 
    $\ln(10^{10} A_s)$ & $\mathcal{U}(1.61,3.91)$ \\
    \hline
    $\Omega_m$ & $\mathcal{U}(0.20,0.40)$ \\
    \hline
    $H_0$ [km/s/Mpc] & $\mathcal{U}(60.0,80.0)$ \\
    \hline
\end{tabular}

%% file: Tables/cfixed.tex
\begin{tabular}{ || c || c | }
    \hline
    Parameter & Value \\
    \hline 
    $\Omega_b h^2$ & $0.02242$ \\
    \hline
    $n_s$ & $0.9665$ \\
    \hline
    $M_\nu$ & $0.06$ eV \\
    \hline
\end{tabular}

%% file: Tables/priors2.tex
\begin{tabular}{ || c || c | }
    \hline
    Parameter & Prior \\
    \hline
    $(1 + b_1)\sigma_8$ & $\mathcal{U}(0.5,3.0)$ \\
    $b_2$ & $\mathcal{N}(0,10)$ \\
    $b_s$ & $\mathcal{N}(0,5)$ \\
    $\alpha_0$ [h$^{-2}$ Mpc$^2$] & $\mathcal{N}(0,30)$ \\
    $\alpha_2$ [h$^{-2}$ Mpc$^2$] & $\mathcal{N}(0,50)$ \\
    $\alpha_x$ [h$^{-2}$ Mpc$^2$] & $\mathcal{N}(0,30)$ \\
    $R_h^3$ [h$^{-3}$ Mpc$^3$] & $\mathcal{N}(0,\frac{1}{3\bar{n}})$ \\
    $R_h^3 \sigma^2$ [h$^{-5}$ Mpc$^5$] & $\mathcal{N}(0,5 \times 10^4)$ \\
    \hline
\end{tabular}

%% file: Tables/priors.tex
\begin{tabular}{ || c || c | }
    \hline
    Parameter & Prior \\
    \hline
    $B_1$ & $\mathcal{U}(0,5.0)$ \\
    $F$ & $\mathcal{U}(0,5.0)$ \\
    $a_{\ell,0}$ & $\mathcal{N}(0,0.05)$ \\
    $a_{\ell,1}$ [$h^{-1}$ Mpc]  & $\mathcal{N}(0,5)$ \\
    \hline
\end{tabular}

%% file: Tables/mock_results.tex
\begin{tabular}{|| c || c | c ||}
    \hline
    & $P_\ell$, $\xi^{\rm rec}_\ell$ & $P_\ell$, $\xi^{\rm rec}_\ell$, $C^{\kappa g}_\ell$ \\
    \hline 
    $\ln(10^{10} A_s)$ & $3.12\pm0.15$ & $3.15\pm0.13$  \\
    $\Omega_m$ & $0.282\pm0.012$ & $0.284\pm0.011$ \\
    $H_0$ [km/s/Mpc] & $69.5\pm1.0$ & $69.5\pm1.0$  \\
    \hline
    $\sigma_8$ & $0.825\pm0.056$ & $0.838\pm0.046$  \\
    \hline
\end{tabular}

%% file: Tables/results.tex
\begin{tabular}{|| c || c | c | c ||}
    \hline
    & $P_\ell$, $\xi^{\rm rec}_\ell$ & $P_\ell$, $\xi^{\rm rec}_\ell$, $C^{\kappa g}_\ell$ & Planck\\
    \hline 
    $\ln(10^{10} A_s)$ & $2.83\pm0.11$ & $2.75\pm0.11$ & $3.044 \pm 0.014$   \\
    $\Omega_m$ & $0.3032\pm0.0084$ & $0.3001\pm0.0078$ &  $0.3153 \pm 0.0073$ \\
    $H_0$ [km/s/Mpc] & $69.21\pm0.78$ & $69.21\pm0.77$ & $67.36 \pm 0.54$  \\
    \hline
    $\sigma_8$ & $0.743\pm0.043$ & $0.707\pm0.035$ &  $0.8111 \pm 0.0060$  \\
    \hline
\end{tabular}

%% file: Tables/results_ns.tex
\begin{tabular}{|| c || c | c | c ||}
    \hline
    & $P_\ell$, $\xi^{\rm rec}_\ell$ & $P_\ell$, $\xi^{\rm rec}_\ell$, $C^{\kappa g}_\ell$ & Planck\\
    \hline 
    $\ln(10^{10} A_s)$ & $2.66\pm 0.16$ & $2.60\pm0.15$ & $3.044 \pm 0.014$   \\
    $\Omega_m$ & $0.322^{+0.015}_{-0.018}$ & $0.316^{+0.014}_{-0.015}$ &  $0.3153 \pm 0.0073$ \\
    $H_0$ [km/s/Mpc] & $70.2\pm1.0$ & $70.0\pm1.0$ & $67.36 \pm 0.54$  \\
    $n_s$ & $0.878\pm 0.060$ & $0.893\pm 0.055$ & $0.9649 \pm 0.0042$\\
    \hline
    $\sigma_8$ & $0.705\pm 0.049$ & $0.674\pm 0.042$ &  $0.8111 \pm 0.0060$  \\
    \hline
\end{tabular}